\documentclass[aps,prapplied,reprint,amsmath,amssymb, floatfix,superscriptaddress,nofootinbib,
longbibliography]{revtex4-2}

\usepackage{graphicx}
\usepackage{subcaption}
\usepackage{booktabs}
\usepackage{multirow}

\usepackage{bm}
\usepackage{quantikz}

\graphicspath{{Plots/}}

\usepackage[
    colorlinks=true,
    linkcolor=blue,
    citecolor=blue,
    urlcolor=blue
]{hyperref}

\usepackage{xcolor}

\begin{document}

\title{Variational Quantum Linear Solver via Block
Encoding for the Poisson Equation}

\author{Viraj Dsouza}
\email[Contact author:]{virajdanieldsouza@gmail.com}
\affiliation{BosonQ Psi Corp., Syracuse, New York 13202, USA}

\author{Ayush Singhal}
\email[]{aysinghal06@gmail.com}
\affiliation{BosonQ Psi Corp., Syracuse, New York 13202, USA}
\affiliation{Department of Computer Science, University of Maryland, College Park, Maryland 20742, USA}

\author{Alex Khan}
\email[]{alex.khan@bqpsim.com}
\affiliation{BosonQ Psi Corp., Syracuse, New York 13202, USA}

\author{Rut Lineswala}
\affiliation{BosonQ Psi Corp., Syracuse, New York 13202, USA}

\author{Abhishek Chopra}
\affiliation{BosonQ Psi Corp., Syracuse, New York 13202, USA}

\begin{abstract}
We present a variational
quantum linear solver (VQLS) for the Poisson equation built on an exact block encoding of the
discrete Laplacian, and demonstrate its performance on
physically motivated benchmarks.
Unlike LCU-based VQLS where the number of distinct circuits required per cost-function
evaluation is $\mathcal{O}(L^2)$, where $L$ is
the number of terms in the LCU decomposition of the discrete Laplacian operator, this approach requires only a single circuit for cost evaluation. We further empirically demonstrate that the choice of classical optimizer materially
affects where the variational optimization ceases to make progress. The solver is benchmarked on three problems: a Poisson equation with
sinusoidal forcing and a steady-state heat conduction problem with a
localized Gaussian source,  both with Dirichlet boundaries, and the pressure-Poisson equation of a
two-dimensional lid-driven cavity flow, in which the solver is invoked
once per time step under Neumann boundary conditions. 
\end{abstract}

\maketitle

%% ============================================================
%% SECTION 1: INTRODUCTION
%% ============================================================

\section{Introduction}
\label{sec:intro}

Solving large linear systems arising from the discretization of
elliptic partial differential equations is one of the most persistent computational bottlenecks in scientific computing, and among these the Poisson equation is particularly ubiquitous.  It governs electrostatic and gravitational potentials, steady-state diffusion and heat conduction, and the pressure field of incompressible flows, and it also arises as an
inner subproblem in a wide range of larger simulation workflows.

Discretizing the equation on a grid with $N$ unknowns often yields a large,
sparse, and structured linear system $A\bm{x} = \bm{b}$, which in
many applications must be solved repeatedly, at every time step of a
transient simulation, for instance.  The cumulative cost of these
elliptic solves can therefore constitute a significant fraction of the total simulation cost.  Computational fluid dynamics (CFD) provides a
demanding and representative instance: incompressible flow solvers built on
projection methods solve an elliptic pressure system at every time step,
and the burden of high-fidelity flow simulation is severe enough that
community roadmaps have identified quantum computing as a long-term
research direction worth
pursuing~\cite{cary2021,mani2023}.

The theoretical foundation for quantum linear solving was laid by
the Harrow--Hassidim--Lloyd (HHL) algorithm~\cite{hhl2009}, which
establishes the possibility of an exponential speedup over classical methods for sparse,
well-conditioned systems under idealized assumptions on sparsity,
condition number, and data access. While HHL has been demonstrated experimentally on small systems~\cite{zheng2017}, practical implementations remain challenging because it requires deep circuits and fault-tolerant hardware~\cite{ibmroadmap,quantinuumroadmap,ionqroadmap}. Similar
limitations apply to the Quantum Singular Value Transformation
(QSVT)~\cite{gilyen2019qsvt}, which provides a unified framework for
matrix inversion, Hamiltonian simulation, and related linear algebra
operations~\cite{martyn2021grand}.  For the near-term Noisy
Intermediate-Scale Quantum (NISQ) era~\cite{preskill_2018}, the
Variational Quantum Linear Solver (VQLS)~\cite{bravoprieto2023} has
emerged as the most practically accessible alternative.  VQLS recasts
$A\bm{x} = \bm{b}$ as a variational optimization problem: a classical
optimizer iteratively updates the parameters $\bm{\theta}$ of a
parametrized quantum circuit $V(\bm{\theta})$ so as to minimize a cost
function quantifying the mismatch between $A\ket{x(\bm{\theta})}$ and
$\ket{b}$.  The approach requires only shallow parametrized circuits and
classical post-processing, and it has been explored for several elliptic
systems including heat conduction~\cite{liu2022,dsouza2025} and the
Poisson equation~\cite{cao2013}.

Two obstacles have limited the practical reach of VQLS on
discretized PDE systems.  The first is the cost of encoding the operator.
In the standard formulation~\cite{bravoprieto2023}, $A$ must be expressed
as a linear combination of $L$ unitaries (LCU), and the cost function is
estimated from $\mathcal{O}(L^2)$ independent Hadamard-test circuits.
For the two-dimensional finite-difference Laplacian with a Pauli-LCU~\cite{koska2024treeapproachpaulidecompositionalgorithm}, the number of circuits required per cost-function evaluation grows 
exponentially in the qubit count $n = \log_2 N$ (Section~\ref{subsec:resources}).  Thus, whatever compression
the quantum state representation provides is forfeited at the level of circuit count alone.  The second obstacle is
trainability. VQLS optimization is affected by barren
plateaus~\cite{mcclean2018}, in which the variance of the cost-function
gradient decays exponentially with number of qubits, making it exponentially
unlikely to sample a parameter configuration with a non-negligible
gradient~\cite{mcclean2018,arrasmith2022equivalence}.  In practice this
manifests as an optimizer that appears to make no progress despite
continued circuit evaluations.

A further and more fundamental caveat applies to quantum linear solvers
generally: the solution is produced as a quantum state $\ket{x}$, and
extracting the full classical solution vector requires $\mathcal{O}(N)$
measurements, which eliminates any potential exponential
speedup~\cite{Aaronson2015ReadFinePrint}.  Practical advantage therefore
requires either that the workflow be restructured to query
low-dimensional functionals of the solution like forces, fluxes, energy
norms, rather than the full field, or that the linear solver be
embedded in a larger quantum pipeline that never requires classical
readout.  Both directions remain open, and the present work does not
resolve them. 

The block encoding~\cite{gilyen2019qsvt} technique offers a route around the first obstacle. Efficient constructions for structured and sparse matrices have been explored~\cite{camps2024,sunderhauf2024,boutot2026}, but their
consequences for VQLS have not been systematically characterized. This paper develops block-encoded VQLS as a solver in its own right and
characterizes it empirically. The block encoding is the unified construction
for discrete Laplacian operators,
a companion contribution by the authors~\cite{boutot2026}, which supports Dirichlet,
Neumann, Robin and periodic boundary conditions across arbitrary spatial
dimensions within a single modular circuit, achieving an exact
$(1, m, 0)$-block encoding with $m = 3 + \lceil\log_2 D\rceil$ ancilla
qubits in $D$ dimensions and gate complexity
$\mathcal{O}(\log N \cdot \log D)$.  

The second obstacle, trainability, is addressed empirically in this work.  We characterize the
barren plateau directly by measuring the decay of
$\mathrm{Var}_{\bm{\theta}}[\partial_\mu C_G]$ with qubit number, and
then examine the Covariance Matrix Adaptation Evolution Strategy
(CMA-ES)~\cite{cma_2001,cma_2006} as an alternative to COBYLA.  CMA-ES
is a population-based, derivative-free optimizer that adapts the full
covariance matrix of its search distribution, enabling it to locate and
follow narrow descent directions in high-dimensional flat landscapes.
Across system sizes from $N = 4$ to $N = 1024$ we find that it delays
the onset of stagnation and remains effective in regimes where COBYLA
makes no progress, and that the parameter regions it visits carry
gradient variances substantially above those of uniformly sampled
parameters.

Our VQLS methodology is assessed on three benchmarks.  The first is a Poisson
problem with smooth sinusoidal forcing.  The second
is a steady-state heat conduction problem with a localized Gaussian
source, modelling a heat-generating component embedded in a conducting
plate. Both carry Dirichlet boundary conditions and are solved in isolation;
the sinusoidal case admits an analytical solution and therefore
provides a controlled accuracy check. The third places the solver inside a projection-method
incompressible Navier--Stokes loop for the two-dimensional lid-driven
cavity, where it is invoked once per time step on a Neumann system whose
right-hand side changes as the flow evolves. This last case tests whether the attainable accuracy
suffices to sustain a time-advancing simulation.  All three are
classically tractable and therefore furnish references against
which the quantum solution can be measured.

The remainder of this paper is structured as follows.
Section~\ref{sec:theory} presents the discretized 2D Poisson problem for
Dirichlet and Neumann boundary conditions, develops the corresponding
block-encoding circuits, and sets out the VQLS methodology.
Section~\ref{sec:results} reports the numerical results: the exactness and post-selection success probability of the encoding, circuit cost relative to the Pauli-LCU VQLS formulation,
solution accuracy on the three benchmark problems, followed by characterization
of the barren plateau and the effect of optimizer choice.
Section~\ref{sec:conclusion} summarises the findings and outlines
directions for future work.

%% ============================================================
%% SECTION 2: THEORY AND METHODOLOGY
%% ============================================================
\section{VQLS via Block Encoding: Theory and Methodology}
\label{sec:theory}
\subsection{The 2D Pressure-Poisson Problem}
\label{subsec:poisson}

In incompressible CFD, projection and fractional-step methods~\cite{chorin1968, guermond2006} enforce the
divergence-free condition on the velocity field by solving a pressure-Poisson
equation of the form
\begin{equation}
  \nabla^2 p = \frac{\rho}{\Delta t} \nabla \cdot \mathbf{u}^*,
  \label{eq:projection}
\end{equation}
where $\mathbf{u}^*$ is the intermediate (non-solenoidal) velocity field,
$\rho$ is the fluid density (taken as unity throughout), and $\Delta t$ is the time step.  The
right-hand side represents the local divergence error introduced during the
velocity prediction step, and solving~\eqref{eq:projection} yields the
pressure correction that restores incompressibility~\cite{gresho1987}.  This elliptic
subproblem is one of the dominant computational bottlenecks in many CFD
workflows, and it is therefore a natural and well-studied benchmark for
linear solvers.

For the first two benchmark problems, we isolate
the elliptic subproblem and consider the general two-dimensional
Poisson equation on the unit square with homogeneous Dirichlet
boundary conditions,
\begin{equation}
  -\nabla^2 u(x,y) = f(x,y), \quad (x,y) \in [0,1]^2,
  \qquad u\big|_{\partial\Omega} = 0,
  \label{eq:poisson}
\end{equation}
where $u$ is the unknown scalar field and $f$ is a prescribed source
term.  Depending on the physical application, $u$ may represent a
pressure field arising from an incompressible flow projection step or
a temperature field arising from steady-state heat conduction.  The
specific source terms used to benchmark the VQLS solver in this work
are described in Section~\ref{sec:results}.

The unit square is discretized on a uniform Cartesian grid with $N_x$
interior points along $x$ and $N_y$ interior points along $y$, with
uniform mesh spacings $\Delta x = 1/(N_x+1)$ and $\Delta y = 1/(N_y+1)$. Dirichlet boundary conditions are imposed on all four domain boundaries. As a result, only the $N = N_x \times N_y$ interior grid points are treated as unknowns.
Applying a standard second-order central finite-difference stencil to
the interior nodes yields the linear system
\begin{equation}
  A\,\mathbf{x} = \mathbf{b},
  \label{eq:linsys}
\end{equation}
where $\mathbf{x} \in \mathbb{R}^N$ contains the interior field
unknowns, and $\mathbf{b} \in \mathbb{R}^N$ contains the discretized
forcing term together with any boundary contributions arising from the
finite-difference stencil. The system matrix $A \in \mathbb{R}^{N \times N}$ inherits the
separable Kronecker-sum structure of the Laplacian operator,
\begin{equation}
  A = I_{N_y} \otimes T_{N_x} + T_{N_y} \otimes I_{N_x},
  \label{eq:kronecker}
\end{equation}
where $I_{N_d}$ denotes the $N_d \times N_d$ identity matrix and
$T_{N_d} \in \mathbb{R}^{N_d \times N_d}$ is the standard
one-dimensional second-order finite-difference Laplacian,
\begin{equation}
  T_{N_d} = \frac{1}{(\Delta x_d)^2}
  \begin{pmatrix}
    -2 &  1 &        &        \\
     1 & -2 &  1     &        \\
       & \ddots & \ddots & \ddots \\
       &        &  1     & -2
  \end{pmatrix} \in \mathbb{R}^{N_d \times N_d}.
  \label{eq:1d_laplacian}
\end{equation}

Here $\Delta x_d= \{\Delta x, \Delta y \}$ based on the $d$ involved. The matrix $A$ is sparse and symmetric. Its sparsity pattern reflects
the standard five-point finite-difference stencil connecting each
interior node to its four nearest neighbours. In the present
implementation, the negative sign associated with the continuous Poisson
operator is absorbed into the right-hand side vector.

We also consider the pressure-Poisson equation~\eqref{eq:projection} posed in a
closed domain with impermeable walls, as in the lid-driven cavity
considered in Section~\ref{sec:results}. For the Neumann case the unknowns are placed at cell centres,
$x_i = (i - \tfrac{1}{2})\Delta x$ with $\Delta x = 1/N_x$, rather than at
the interior vertices of the Dirichlet grid.  Imposing
$\partial p/\partial n = 0$ through the ghost-node relation
$p_{-1} = p_{0}$ modifies only the first and last rows of each
one-dimensional factor, leaving the Kronecker-sum structure
of~\eqref{eq:kronecker} intact,
\begin{equation}
  A^{\mathrm{N}}
  = I_{N_y} \otimes T^{\mathrm{N}}_{N_x}
  + T^{\mathrm{N}}_{N_y} \otimes I_{N_x},
  \label{eq:kronecker_neumann}
\end{equation}
with the one-dimensional Neumann operator
\begin{equation}
  T^{\mathrm{N}}_{N_d} = \frac{1}{(\Delta x_d)^2}
  \begin{pmatrix}
    -1 &  1 &        &        \\
     1 & -2 &  1     &        \\
       & \ddots & \ddots & \ddots \\
       &        &  1     & -1
  \end{pmatrix}.
  \label{eq:1d_laplacian_neumann}
\end{equation}
Comparing~\eqref{eq:1d_laplacian_neumann} with~\eqref{eq:1d_laplacian},
the sole difference is the diagonal entry at the two boundary nodes,
reduced from $-2$ to $-1$. Two
properties of $A^{\mathrm{N}}$ (here the superscript ${\mathrm{N}}$ is a label for Neumann)  bear on the quantum solve.  It is singular,
with null space spanned by the constant vector $\mathbf{1}$, so the
solution is determined only up to an additive constant; we fix this gauge
by projecting the recovered field onto the zero-mean subspace, which
leaves $\nabla p$ and hence the corrected velocity unchanged. A solution exists provided the data are compatible,
$\mathbf{1}^{\!\top}\bm{b} = 0$, which holds for the pressure system provided the normal velocity
component of $\mathbf{u}^*$ vanishes on every wall face entering the
discrete divergence. ~\cite{gresho1987}

Both operators have spectral norm growing as
$\mathcal{O}(1/h_{\min}^2)$ with $h_{\min} = \min(\Delta x, \Delta y)$,
whereas block encoding requires a matrix of spectral norm at most unity.
The system must therefore be rescaled.  The particular scaling used here is dictated by the structure
of the encoding and is introduced in
Section~\ref{subsec:block_encoding}.

We consider a range of proof-of-concept discretization sizes spanning
from $N=4$ to $N=1024$ interior unknowns, corresponding to uniform grids
ranging from $2 \times 2$ to $32 \times 32$ interior
points.  These problem sizes are intentionally modest and are chosen to
remain compatible with the qubit and circuit-depth limitations of
current quantum simulators.  In all cases, the
classical reference solution is obtained using the
Successive Over-Relaxation (SOR) iterative method, against which the VQLS
solution is compared to assess solution accuracy.

\subsection{Block Encoding of the 2D Laplacian}
\label{subsec:block_encoding}

Block encoding provides the standard mechanism for embedding a
generally non-unitary matrix into a larger unitary operator that
can be realised as a quantum circuit. Formally, a unitary
$U_A \in \mathbb{C}^{2^{n+m} \times 2^{n+m}}$ acting on $m$
ancilla qubits and $n$ system qubits is called an
$(\alpha, m, \varepsilon)$-block encoding of $A \in
\mathbb{C}^{2^n \times 2^n}$ if
\begin{equation}
  \left\|A - \alpha
    \bigl(\langle 0|^{\otimes m} \otimes I\bigr)\,
    U_A\,
    \bigl(|0\rangle^{\otimes m} \otimes I\bigr)
  \right\|_2 \leq \varepsilon,
  \label{eq:be_def}
\end{equation}
where $\alpha > 0$ is the sub-normalization factor and $\varepsilon \geq 0$
is the approximation error.  When $\varepsilon = 0$ the encoding is said
to be \emph{exact}.  In block-matrix form, $U_A$ takes the structure
\begin{equation}
  U_A = \begin{pmatrix} A/\alpha & * \\ * & * \end{pmatrix},
  \label{eq:be_matrix}
\end{equation}
where the asterisks denote irrelevant blocks.  Post-selecting the ancilla
register on the outcome $|0\rangle^{\otimes m}$ projects the system
register onto a state proportional to $A|\psi\rangle$, with success
probability $\|A|\psi\rangle\|^2/\alpha^2$.

The block-encoding requirement $\|A\|_2 \leq 1$ is not satisfied by
the system matrix $A$ introduced in~\eqref{eq:kronecker}, whose
spectral norm grows as $\mathcal{O}(1/h_{\min}^2)$.  We therefore
introduce the globally scaled 2D Dirichlet Laplacian
\begin{equation}
  \widetilde{A} := \frac{\Lambda}{4}\, A,
  \qquad
  \Lambda := \left(\frac{1}{\Delta x^2} + \frac{1}{\Delta y^2}\right)^{-1},
  \label{eq:scaled_2d}
\end{equation}
which satisfies $\|\widetilde{A}\|_2 \leq 1$ by construction.
The Kronecker-sum structure of~\eqref{eq:kronecker} is preserved
under this scaling, giving
\begin{equation}
  \widetilde{A}
  = \omega_1\,\widetilde{A}^{(N_x)} \otimes I_{N_y}
  + \omega_2\, I_{N_x} \otimes \widetilde{A}^{(N_y)},
  \label{eq:2d_scaled}
\end{equation}
where $\widetilde{A}^{(N_d)} := \frac{\Delta x_d^2}{4} T_{N_d}$
is the scaled 1D component along dimension $d$, satisfying
$\|\widetilde{A}^{(N_d)}\|_2 \leq 1$, and the dimension-dependent
weights
\begin{equation}
  \omega_1 = \frac{1/\Delta x^2}{1/\Delta x^2 + 1/\Delta y^2},
  \qquad
  \omega_2 = 1 - \omega_1,
  \label{eq:weights}
\end{equation}
satisfy $\omega_1 + \omega_2 = 1$.  The linear system to be solved
by VQLS is accordingly $\widetilde{A}\,\mathbf{x} = \widetilde{\mathbf{b}}$,
where $\widetilde{\mathbf{b}} = (\Lambda/4)\,\mathbf{b}$.  For a uniform grid, $\Delta x = \Delta y$, hence we get $\omega_1 = \omega_2 = 1/2$.

We employ the unified construction of~\cite{boutot2026} to build an
explicit quantum circuit that block encodes $\widetilde{A}$.  The
construction exploits the Kronecker-sum separability
of~\eqref{eq:2d_scaled} by introducing a one-qubit \emph{selector}
register $|k\rangle$ that is prepared in the superposition
\begin{equation}
  U_{\mathrm{prep}_k}|0\rangle
  = \sqrt{\omega_1}|0\rangle + \sqrt{\omega_2}|1\rangle,
  \label{eq:state_prep}
\end{equation}
implemented by a single $R_y(\theta_0)$ rotation with
$\theta_0 = 2\arccos\!\sqrt{\omega_1}$. Conditioned on
$|k\rangle = |d\rangle$ ($d \in \{1,2\}$), a one-dimensional
block-encoding sub-circuit $U_{d}$ for the boundary condition imposed
along axis $d$ is applied to the system register $|j^{(d)}\rangle$,
after which $U^\dagger_{\mathrm{prep}_k}$ uncomputes the selector.
This selector construction is independent of the boundary condition;
only the one-dimensional sub-circuits differ.  We describe the
Dirichlet and Neumann sub-circuits in turn, and we note that the same
scaling $\|\widetilde{A}^{(N_d)}\|_2 \leq 1$ holds in both cases, since
the spectrum of $T^{\mathrm{N}}_{N_d}$ likewise lies in
$[-4/\Delta x_d^2,\,0]$.

\subsubsection{Dirichlet boundary conditions}
\label{subsubsec:be_dirichlet}

Each 1D sub-circuit $U_{d}$ is itself
constructed from cyclic shift operators $S^-|j\rangle =
|j-1\bmod N_d\rangle$ and $S^+|j\rangle = |j+1\bmod N_d\rangle$,
a Hadamard--$Z$--Hadamard ladder on two ancilla qubits
$|\ell_0\rangle$, $|\ell_1\rangle$, and one additional
$|\mathrm{del}\rangle$ ancilla that suppresses the unwanted
wrap-around coupling at the Dirichlet boundaries $j=0$ and
$j=N_d-1$; see~\cite{boutot2026} for the full construction and proof.

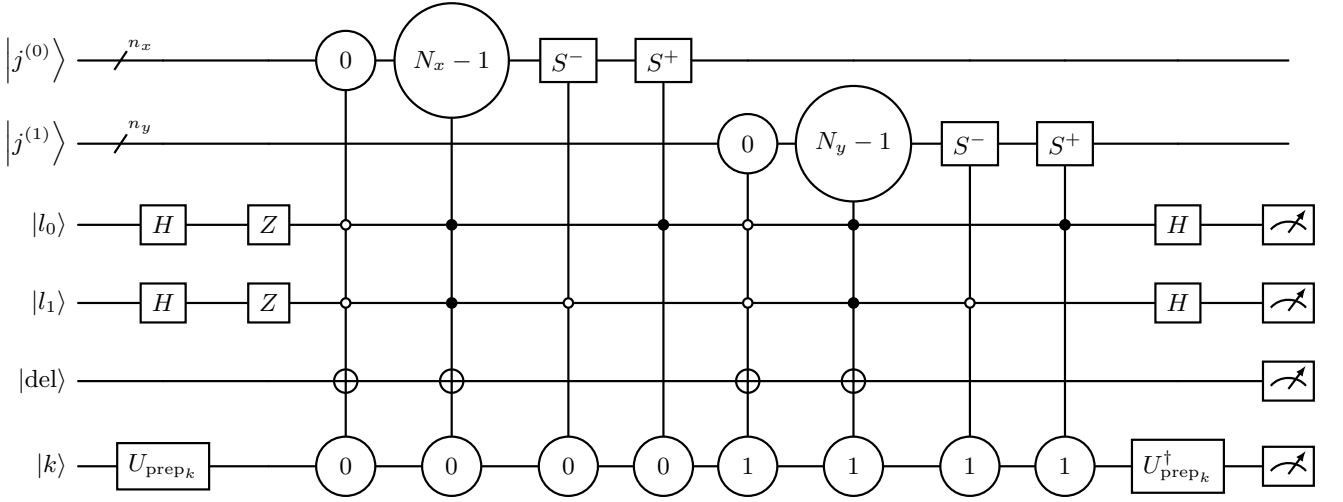
\begin{figure*}[t]
\centering
\resizebox{0.98\textwidth}{!}{%
\begin{quantikz}
    \lstick{$\ket{j^{(0)}}$} & \qwbundle{n_x} & & \gate[style={draw, circle, fill=white, inner sep=1pt}]{0}\wire[d]{q} & \gate[style={draw, circle, fill=white, inner sep=1pt}]{N_x-1}\wire[d]{q} & \gate{S^-} & \gate{S^+} & & & & & & \\
    \lstick{$\ket{j^{(1)}}$} & \qwbundle{n_y} & & \wire[d]{q} & \wire[d]{q} & & & \gate[style={draw, circle, fill=white, inner sep=1pt}]{0}\wire[d]{q} & \gate[style={draw, circle, fill=white, inner sep=1pt}]{N_y-1}\wire[d]{q} & \gate{S^-} & \gate{S^+} & & \\
    \lstick{$\ket{l_0}$} & \gate{H} & \gate{Z} & \ctrl[open]{1} & \ctrl{1} & & \ctrl{-2} \wire[d]{q} & \ctrl[open]{1} & \ctrl{1} & & \ctrl{-1}\wire[d]{q} & \gate{H} & \meter{} \\
    \lstick{$\ket{l_1}$} & \gate{H} & \gate{Z} & \ctrl[open]{1} & \ctrl{1} & \ctrl[open]{-3}\wire[d]{q} & \wire[d]{q} & \ctrl[open]{1} & \ctrl{1} & \ctrl[open]{-2}\wire[d]{q} & \wire[d]{q} & \gate{H} & \meter{} \\
    \lstick{$\ket{\mathrm{del}}$} & & & \targ{} & \targ{} & & & \targ{} & \targ{} & & & & \meter{} \\
    \lstick{$\ket{k}$} & \gate{U_{\mathrm{prep}_k}} & & \gate[style={draw, circle, fill=white, inner sep=1pt}]{0}\wire[u]{q} & \gate[style={draw, circle, fill=white, inner sep=1pt}]{0}\wire[u]{q} & \gate[style={draw, circle, fill=white, inner sep=1pt}]{0}\wire[u]{q} & \gate[style={draw, circle, fill=white, inner sep=1pt}]{0}\wire[u]{q} & \gate[style={draw, circle, fill=white, inner sep=1pt}]{1}\wire[u]{q} & \gate[style={draw, circle, fill=white, inner sep=1pt}]{1}\wire[u]{q} & \gate[style={draw, circle, fill=white, inner sep=1pt}]{1}\wire[u]{q} & \gate[style={draw, circle, fill=white, inner sep=1pt}]{1}\wire[u]{q} & \gate{U_{\mathrm{prep}_k}^\dagger} & \meter{} \\
\end{quantikz}
}
\caption{Block encoding circuit for the 2D Laplacian with Dirichlet boundary
conditions along both spatial directions, following the construction
of~\cite{boutot2026}. Post-selecting all four ancilla qubits on $|0\rangle^{\otimes 4}$ realises
the action of $\widetilde{A}$ on the system register.}
\label{fig:be_circuit}
\end{figure*}

The complete circuit, shown in Figure~\ref{fig:be_circuit}, uses
$m = 4$ ancilla qubits in total and constitutes an exact
$(1, 4, 0)$-block encoding of $\widetilde{A}$.

Following the analytical resource estimates of~\cite{boutot2026}, the
T-gate count of the 2D Dirichlet block-encoding circuit scales as
$\mathcal{O}(\log N \cdot \log D)$, which for fixed dimension $D=2$
reduces to $\mathcal{O}(\log N)$. This poly-logarithmic scaling in
the matrix dimension is the central efficiency property of the
construction.

\subsubsection{Neumann boundary conditions}
\label{subsubsec:be_neumann}

\begin{figure*}[t]
\centering
\begin{subfigure}[b]{0.52\textwidth}
  \centering
  \resizebox{\textwidth}{!}{%
  \begin{quantikz}
      \lstick{$\ket{j}$} & \qwbundle{n} & &
      \gate[style={draw, circle, fill=white, inner sep=1pt}]{0}\wire[d]{q} &
      \gate[style={draw, circle, fill=white, inner sep=1pt}]{0}\wire[d]{q} &
      \gate[style={draw, circle, fill=white, inner sep=1pt}]{N-1}\wire[d]{q} &
      \gate[style={draw, circle, fill=white, inner sep=1pt}]{N-1}\wire[d]{q} &
      \gate{S^-} & \gate{S^+} & & \\
    \lstick{$\ket{\ell_0}$} & \gate{H} & \gate{Z} &
      \ctrl[open]{1} & \ctrl{1} & \ctrl{1} & \ctrl{1} &
      & \ctrl{-1} & \gate{H} & \meter{} \\
    \lstick{$\ket{\ell_1}$} & \gate{H} & \gate{Z} &
      \ctrl[open]{1} & \ctrl[open]{1} & \ctrl{1} & \ctrl[open]{1} &
      \ctrl[open]{-2} & & \gate{H} & \meter{} \\
    \lstick{$\ket{\mathrm{del}}$} & & &
      \targ{} & \targ{} & \targ{} & \targ{} & & & & \meter{}
  \end{quantikz}}
  \caption{One-dimensional Neumann sub-circuit $U^{\mathrm{N}}_d$.}
  \label{fig:be_1d_neumann}
\end{subfigure}
\hfill
\begin{subfigure}[b]{0.46\textwidth}
  \centering
  \resizebox{\textwidth}{!}{%
  \begin{quantikz}
      \lstick{$\ket{j^{(0)}}$} & \qwbundle{n_x} & &
      \gate[2]{U^{\mathrm{N}}_{1}} & & & \\
    \lstick{$\ket{\ell_0 \ell_1\, \mathrm{del}}$} & \qwbundle{3} & & &
      \gate[2]{U^{\mathrm{N}}_{2}} & & \meter{} \\
    \lstick{$\ket{j^{(1)}}$} & \qwbundle{n_y} & & & & & \\
    \lstick{$\ket{k}$} & & \gate{U_{\mathrm{prep}_k}} &
      \ctrl[open]{-3} & \ctrl{-2} &
      \gate{U^{\dagger}_{\mathrm{prep}_k}} & \meter{}
  \end{quantikz}}
  \caption{2D Laplacian with Neumann boundary
conditions along both spatial directions}
  \label{fig:be_2d_neumann}
\end{subfigure}
\caption{Block-encoding circuits for the Laplacian with Neumann
boundary conditions~\cite{boutot2026}.  In (b), $U^{\mathrm{N}}_1$ and
$U^{\mathrm{N}}_2$ denote the sub-circuit of (a) applied to the $x$- and
$y$-registers.  Post-selecting all four ancilla qubits on
$|0\rangle^{\otimes 4}$ realises the action of
$\widetilde{A}^{\mathrm{N}}$.}
\label{fig:be_neumann}
\end{figure*}
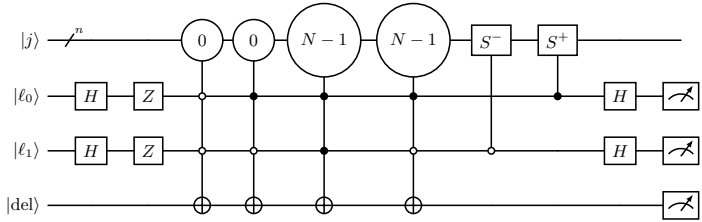
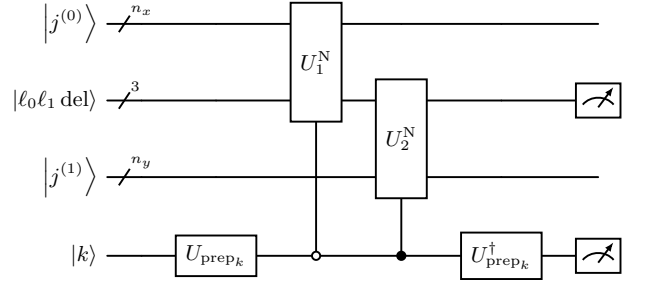

The pressure-Poisson system of
Section~\ref{subsec:poisson} carries homogeneous Neumann
conditions on all four walls, and the corresponding scaled operator
$\widetilde{A}^{\mathrm{N}}$ differs from the Dirichlet operator only in
the boundary rows of each one-dimensional factor. The framework of~\cite{boutot2026} accommodates this case within the
same circuit template. 
Dirichlet closure controls as shown in Figure ~\ref{fig:be_circuit} deletes the two wrap-around contributions of the
cyclic shifts; Neumann closure additionally deletes one of the two
diagonal branches at $j = 0$ and $j = N-1$, which is precisely what
reduces the boundary diagonal from $-2$ to $-1$
in~\eqref{eq:kronecker_neumann}. This costs two further
comparator-controlled $X$ gates, shown in
Figure~\ref{fig:be_1d_neumann}, and the encoding remains exact.
See~\cite{boutot2026} for the full construction and proof.

The two-dimensional Neumann block encoding is assembled by the same
selector construction used in the Dirichlet case.  Writing
$U^{\mathrm{N}}_{1}$ and $U^{\mathrm{N}}_{2}$ for the one-dimensional
sub-circuits of Figure~\ref{fig:be_1d_neumann} acting on the registers
$|j^{(0)}\rangle$ and $|j^{(1)}\rangle$,
Figure~\ref{fig:be_2d_neumann} shows the circuit in compact form.

\subsection{Cost Function Evaluation and Solution Recovery}
\label{subsec:vqls_theory}

Given the scaled linear system $\widetilde{A}\,\mathbf{x} =
\widetilde{\mathbf{b}}$, VQLS seeks a quantum state proportional
to the solution vector by preparing a parametrized ansatz state
\begin{equation}
  |x(\bm{\theta})\rangle = V(\bm{\theta})|0\rangle^{\otimes n},
  \label{eq:ansatz}
\end{equation}
where $V(\bm{\theta})$ is a parametrized unitary circuit acting on
$n$ system qubits and $\bm{\theta} \in \mathbb{R}^p$ is the vector
of variational parameters to be optimised.  The goal is to find
$\bm{\theta}^*$ such that $|x(\bm{\theta}^*)\rangle \propto
\widetilde{A}^{-1}|\widetilde{b}\rangle$, where
$|\widetilde{b}\rangle = U|0\rangle^{\otimes n}$ is the normalized
right-hand side state prepared by a unitary $U$~\cite{M_tt_nen_2004}. To drive the optimization towards this target, one needs a scalar cost function
$C_G(\bm{\theta})$ that is (i) efficiently computable on a quantum
circuit, (ii) equal to zero if and only if
$\widetilde{A}|x(\bm{\theta})\rangle \propto |\widetilde{b}\rangle$,
and (iii) positive otherwise.

A natural candidate is the normalized global cost
function~\cite{bravoprieto2023}
\begin{equation}
  C_G(\bm{\theta})
  = 1 - \frac{
      \left|\langle\widetilde{b}|
        \widetilde{A}|x(\bm{\theta})\rangle
      \right|^2}
    {\langle x(\bm{\theta})|
      \widetilde{A}^\dagger\widetilde{A}
      |x(\bm{\theta})\rangle},
  \label{eq:cost_raw}
\end{equation}
which measures the squared overlap between the normalized vectors
$\widetilde{A}|x(\bm{\theta})\rangle/\|\widetilde{A}|x(\bm{\theta})\rangle\|$
and $|\widetilde{b}\rangle$.  By the Cauchy--Schwarz inequality,
$C_G(\bm{\theta}) \geq 0$ always holds.  Moreover, $C_G(\bm{\theta})
= 0$ if and only if $\widetilde{A}|x(\bm{\theta})\rangle$ is
proportional to $|\widetilde{b}\rangle$, i.e.\ if and only if
$|x(\bm{\theta})\rangle \propto \widetilde{A}^{-1}|\widetilde{b}\rangle$,
which is the desired solution state.  Minimising $C_G(\bm{\theta})$
therefore directly drives the ansatz state towards the solution of
the linear system.  An equivalent and more convenient expression is
obtained by recognising that the denominator is simply
$\|{\widetilde{A}}|x(\bm{\theta})\rangle\|^2$, so that the cost can be written as
\begin{equation}
  C_G(\bm{\theta})
  = 1 -
    \left|\langle\widetilde{b}|
      \frac{\widetilde{A}|x(\bm{\theta})\rangle}
           {\|\widetilde{A}|x(\bm{\theta})\rangle\|}
    \right|^2
  = 1 - \left|\langle 0^n|\Psi_\mathrm{post}(\bm{\theta})\rangle\right|^2,
  \label{eq:cost}
\end{equation}
where the post-selected state $|\Psi_\mathrm{post}(\bm{\theta})\rangle$
is defined in~\eqref{eq:postsel_norm} below.  The cost is therefore
the complement of the fidelity between the normalized state
$\widetilde{A}|x(\bm{\theta})\rangle$ and the target $|\widetilde{b}\rangle$:
it equals zero at the solution and increases monotonically as the
ansatz deviates from it.

In the block-encoding formulation~\cite{dsouza2025},
we assume the $(\alpha, m, 0)$-block-encoding unitary $U_{\widetilde{A}}$
of Section~\ref{subsec:block_encoding}, with $\alpha = 1$ and $m = 4$
ancilla qubits for the discrete 2D Laplacian operator.  The extended Hamiltonian operator is defined as,
\begin{equation}
  H_G^{(\mathrm{ext})} :=
    U_{\widetilde{A}}^\dagger
    \Bigl(|0^m\rangle\langle 0^m|_a
          \otimes
          \bigl(I_s - |\widetilde{b}\rangle\langle\widetilde{b}|_s\bigr)
    \Bigr)
    U_{\widetilde{A}},
  \label{eq:hamiltonian}
\end{equation}
where the subscripts $a$ and $s$ refer to the ancilla and system
subspaces respectively.  For the input state
$|X_\mathrm{ext}\rangle = |0^m\rangle_a \otimes |x(\bm{\theta})\rangle_s$,
the expectation value of this Hamiltonian is
\begin{equation}
  \langle X_\mathrm{ext}|H_G^{(\mathrm{ext})}|X_\mathrm{ext}\rangle
  = \frac{1}{\alpha^2}
    \langle x(\bm{\theta})|
    \widetilde{A}^\dagger
    \bigl(I - |\widetilde{b}\rangle\langle\widetilde{b}|\bigr)
    \widetilde{A}
    |x(\bm{\theta})\rangle,
  \label{eq:expect}
\end{equation}
which corresponds to the numerator of $C_G(\bm{\theta})$ up to the
factor $\alpha^{-2}$.  Evaluating this expectation value on a quantum
circuit proceeds as follows.  One prepares
\begin{equation}
  |\Psi'(\bm{\theta})\rangle
  := U_{\widetilde{A}}\bigl(|0^m\rangle \otimes |x(\bm{\theta})\rangle\bigr),
  \qquad
  U|0^n\rangle = |\widetilde{b}\rangle,
  \label{eq:psi_prime}
\end{equation}
and post-selects the ancilla register on $|0^m\rangle$, yielding the
unnormalized system state
\begin{equation}
  \bigl(\langle 0^m| \otimes U^\dagger\bigr)|\Psi'(\bm{\theta})\rangle
  = \frac{1}{\alpha}\,U^\dagger \widetilde{A}|x(\bm{\theta})\rangle.
  \label{eq:postsel_state}
\end{equation}
The probability of obtaining the ancilla outcome $|0^m\rangle$ is
\begin{equation}
  p(\bm{\theta})
  = \frac{\langle x(\bm{\theta})|
          \widetilde{A}^\dagger \widetilde{A}
          |x(\bm{\theta})\rangle}{\alpha^2},
  \label{eq:prob}
\end{equation}
and the normalized post-selected state is
\begin{equation}
  |\Psi_\mathrm{post}(\bm{\theta})\rangle
  = \frac{U^\dagger \widetilde{A}|x(\bm{\theta})\rangle}
         {\sqrt{\langle x(\bm{\theta})|
                \widetilde{A}^\dagger \widetilde{A}
                |x(\bm{\theta})\rangle}}.
  \label{eq:postsel_norm}
\end{equation}
Substituting into~\eqref{eq:cost} and using Bayes' theorem, the
fidelity term becomes
\begin{equation}
\begin{split}
  \left|\langle 0^n|\Psi_\mathrm{post}(\bm{\theta})\rangle\right|^2
  &= P\!\left(\mathrm{sys}=|0^n\rangle \;\middle|\;
     \mathrm{anc}=|0^m\rangle\right) \\[4pt]
  &= \frac{P\!\left(|0\rangle^{\otimes(n+m)}\right)}
          {P\!\left(|0\rangle^{\otimes m}\right)},
\end{split}
\label{eq:bayes}
\end{equation}
so that the global cost function is entirely determined by two
measurement probabilities obtainable from a single quantum circuit,
and the final expression for the cost is
\begin{equation}
  C_G(\bm{\theta})
  = 1 - \frac{P\!\left(|0\rangle^{\otimes(n+m)}\right)}
             {P\!\left(|0\rangle^{\otimes m}\right)}.
  \label{eq:cost_final}
\end{equation}

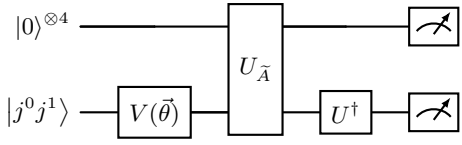
\begin{figure}[htbp]
  \centering
  \begin{quantikz}
  \lstick{$\ket{0}^{\otimes 4}$} & \qw & \gate[wires=2]{U_{\widetilde{A}}} & \qw & \meter{} \\
  \lstick{$\ket{j^0j^1}$} & \gate{V(\vec{\theta})} & \qw & \gate{U^\dagger} & \meter{}
\end{quantikz}

  \caption{Schematic VQLS circuit for the 2D Poisson problem via
    block encoding.}
  \label{fig:vqls_circuit}
\end{figure}

A key advantage of the block-encoding formulation is that
$C_G(\bm{\theta})$ is estimated from the measurement statistics
of a \emph{single} quantum circuit via~\eqref{eq:cost_final},
with $m = 4$ for the 2D Dirichlet/Neumann Laplacian case. Post-selecting the ancilla register on $|0\rangle^{\otimes 4}$ and measuring all
qubits yields the numerator and denominator of
Eq.~\eqref{eq:cost_final}, from which $C_G(\bm{\theta})$ is computed classically.  This contrasts with the LCU based VQLS formulation~\cite{bravoprieto2023}, which requires $\mathcal{O}(L^2)$ separate Hadamard-test circuits for computing the cost function.  The schematic VQLS circuit implementing this evaluation is shown in
Figure~\ref{fig:vqls_circuit}.

The VQLS optimization returns the normalized state
$|x(\bm{\theta}^*)\rangle$ satisfying
$\widetilde{A}|x(\bm{\theta}^*)\rangle \propto |\widetilde{b}\rangle$.
It therefore encodes the direction of the solution but not its
magnitude. Writing
$\mathbf{x} = \|\mathbf{x}\|\,|x(\bm{\theta}^*)\rangle$ and
substituting into~\eqref{eq:linsys} gives
$A|x(\bm{\theta}^*)\rangle = \mathbf{b}/\|\mathbf{x}\|$; taking norms
yields
\begin{equation}
  \|\mathbf{x}\|
  = \frac{\|\mathbf{b}\|}
         {\sqrt{\langle x(\bm{\theta}^*)|A^\dagger A|x(\bm{\theta}^*)\rangle}},
  \label{eq:norm_recovery}
\end{equation}
so that the classical solution vector is reconstructed as
\begin{equation}
  \mathbf{x}
  = \frac{\|\mathbf{b}\|}
         {\sqrt{\langle x(\bm{\theta}^*)|A^\dagger A|x(\bm{\theta}^*)\rangle}}
    \;|x(\bm{\theta}^*)\rangle .
  \label{eq:solution_recovery}
\end{equation}
The quantity in the denominator is obtained coherently, from the
post-selection statistics already collected during cost evaluation:
by~\eqref{eq:prob} and~\eqref{eq:scaled_2d},
$P(|0\rangle^{\otimes 4})
= (\Lambda/4)^2 \langle x(\bm{\theta})|A^\dagger A|x(\bm{\theta})\rangle$,
so that
$\sqrt{\langle x(\bm{\theta}^*)|A^\dagger A|x(\bm{\theta}^*)\rangle}
= (4/\Lambda)\sqrt{P(|0\rangle^{\otimes 4})}$ with $\Lambda$ a known
classical constant.  Estimating it to relative
precision $\varepsilon$ costs $\mathcal{O}(1/(p\varepsilon^2))$ shots
with $p$ the post-selection success probability, independently of $N$,
which is a further practical benefit of the improved success probability
of the construction of~\cite{boutot2026}.

% One remark specific to the singular Neumann system
% of~\eqref{eq:kronecker_neumann} is in order.  The cost
% function~\eqref{eq:cost} is exactly invariant under the addition of any
% component along $\mathbf{1}$ to $|x(\bm{\theta})\rangle$, since such a
% component lies in $\mathrm{null}(A^{\mathrm{N}})$ and cancels between
% the numerator and the denominator.  The optimizer therefore receives no
% signal about it and will not remove it, and the zero-mean projection
% described in Section~\ref{subsec:poisson} must be applied to the
% recovered field.  The rescaling
% in~\eqref{eq:solution_recovery} remains exact under this projection.

\begin{figure}
\centering
\begin{quantikz}
    \lstick{$\ket{q_1}$} & \gate{R_y} & \ctrl{1}   & \qw       & \gate{R_y} & \qw         & \qw         & \qw \\
    \lstick{$\ket{q_2}$} & \gate{R_y} & \ctrl{-1}  & \qw       & \gate{R_y} & \ctrl{1}    & \gate{R_y}  & \qw \\
    \lstick{$\ket{q_3}$} & \gate{R_y} & \ctrl{1}   & \qw       & \gate{R_y} & \ctrl{-1}   & \gate{R_y}  & \qw \\
    \lstick{$\ket{q_4}$} & \gate{R_y} & \ctrl{-1}  & \qw       & \gate{R_y} & \ctrl{1}    & \gate{R_y}  & \qw \\
    \lstick{$\ket{q_5}$} & \gate{R_y} & \ctrl{1}   & \qw       & \gate{R_y} & \ctrl{-1}   & \gate{R_y}  & \qw \\
    \lstick{$\vdots$}    & \vdots     &            &           & \vdots     &             & \vdots      &      \\
    \lstick{$\ket{q_{n-1}}$} & \gate{R_y} & \ctrl{1}   & \qw       & \gate{R_y} & \ctrl{-1}   & \gate{R_y}  & \qw \\
    \lstick{$\ket{q_{n}}$} & \gate{R_y} & \ctrl{-1}  & \qw       & \gate{R_y} & \qw         & \qw         & \qw \\
\end{quantikz}
\caption{Hardware-Efficient ansatz with one layer for $V(\vec{\theta})$.}
\label{fig:VQC}
\end{figure}
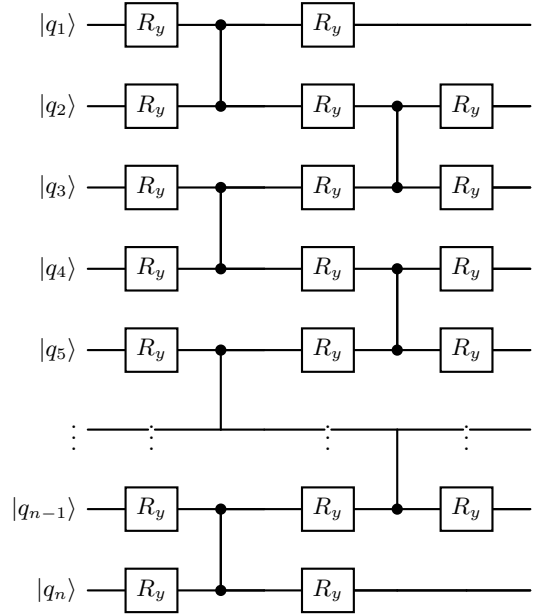

For the variational circuit $V(\bm{\theta})$, we utilize the
hardware-efficient ansatz (HEA)~\cite{Kandala_2017} comprising alternating
single-qubit $R_y$ rotation gates and two-qubit controlled-$Z$
(CZ) entangling gates.  Each layer consists of
$R_y(\theta_i)$ rotations applied to all $n$ system qubits,
followed by CZ gates on alternating pairs of neighbouring qubits,
followed by a second round of $R_y$ rotations, as shown in the Figure~\ref{fig:VQC}.  For an
ansatz with $l$ layers, the total number of variational parameters is
$p = 2l (n-2) + n$, which scales linearly with the number of system qubits.

\subsection{The Barren-Plateau Phenomenon}
\label{subsec:barren}

A challenge shared by all variational quantum algorithms, VQLS included,
is the barren-plateau phenomenon.  As the number of qubits $n$ or the
circuit depth increases, the gradient of the cost function with respect
to the variational parameters becomes exponentially suppressed.
Formally, $C_G(\bm{\theta})$ exhibits a \emph{probabilistic barren
plateau}~\cite{mcclean2018} if, for $\bm{\theta}$ sampled from some
distribution,
\begin{equation}
  \mathrm{Var}_{\bm{\theta}}[\partial_\mu C_G(\bm{\theta})]
  \in \mathcal{O}(b^{-n}),
  \label{eq:barren}
\end{equation}
for some $b > 1$ and some $\theta_\mu \in \bm{\theta}$: as $n$ grows,
the probability of sampling a parameter vector with a non-negligible
gradient becomes exponentially small.  The behaviour is attributed to
the emergence of unitary 2-design characteristics in sufficiently deep
or expressive circuits, and global cost functions of the form
of~\eqref{eq:cost} are known to be particularly
susceptible~\cite{mcclean2018,cerezo2021cost}.

It is important to note, however, that barren plateaus describe an
\emph{average-case} property of the cost landscape rather than a
uniform absence of gradients everywhere. Even in high-dimensional
parameter spaces dominated by flat regions-- narrow
``gorges''~\cite{arrasmith2022equivalence}, localized directions of non-negligible gradient
magnitude can persist around the global minimum, but occupy an exponentially small fraction of the
parameter space. Optimization success
therefore depends less on the typical gradient magnitude than on
whether the optimizer can find and stay within these atypical regions.  Deterministic gradient-free methods relying on local differential information tend to
stall once typical gradients are suppressed; on the contrary, population-based methods
maintain broader exploration and may sample gorge-like directions that
local models never see.  Section~\ref{subsec:optimizer_results} tests
this empirically, measuring~\eqref{eq:barren} both under uniform
sampling and along the trajectories the CMA-ES optimizer follows.

%% SECTION 3: RESULTS
%% ============================================================
\section{Results}
\label{sec:results}

Section~\ref{subsec:resources} characterizes the block-encoding circuit
itself: its exactness, its post-selection success probability, and its
cost relative to the Pauli-LCU formulation of~\cite{bravoprieto2023}.
Section~\ref{subsec:accuracy} then reports solution accuracy on the
three benchmark problems, and
Section~\ref{subsec:optimizer_results} examines the barren plateau and
the effect of optimizer choice.
All simulations are performed on the Qiskit Aer statevector simulator.
Classical reference solutions are obtained with the Successive
Over-Relaxation (SOR) method applied to the identical discrete system.  Because the variational optimization depends on the
random initialization of $\bm{\theta}$, every configuration is run with
five independent seeds; we report the geometric mean together with
$\pm 1$ standard deviation in $\log_{10}$ across seeds.

\subsection{Circuit Verification and Resource Comparison}
\label{subsec:resources}

\subsubsection{Exactness of the encoding}

\begin{figure*}[t]
  \centering
  \begin{subfigure}[t]{0.48\textwidth}
    \centering
    \includegraphics[width=\textwidth]{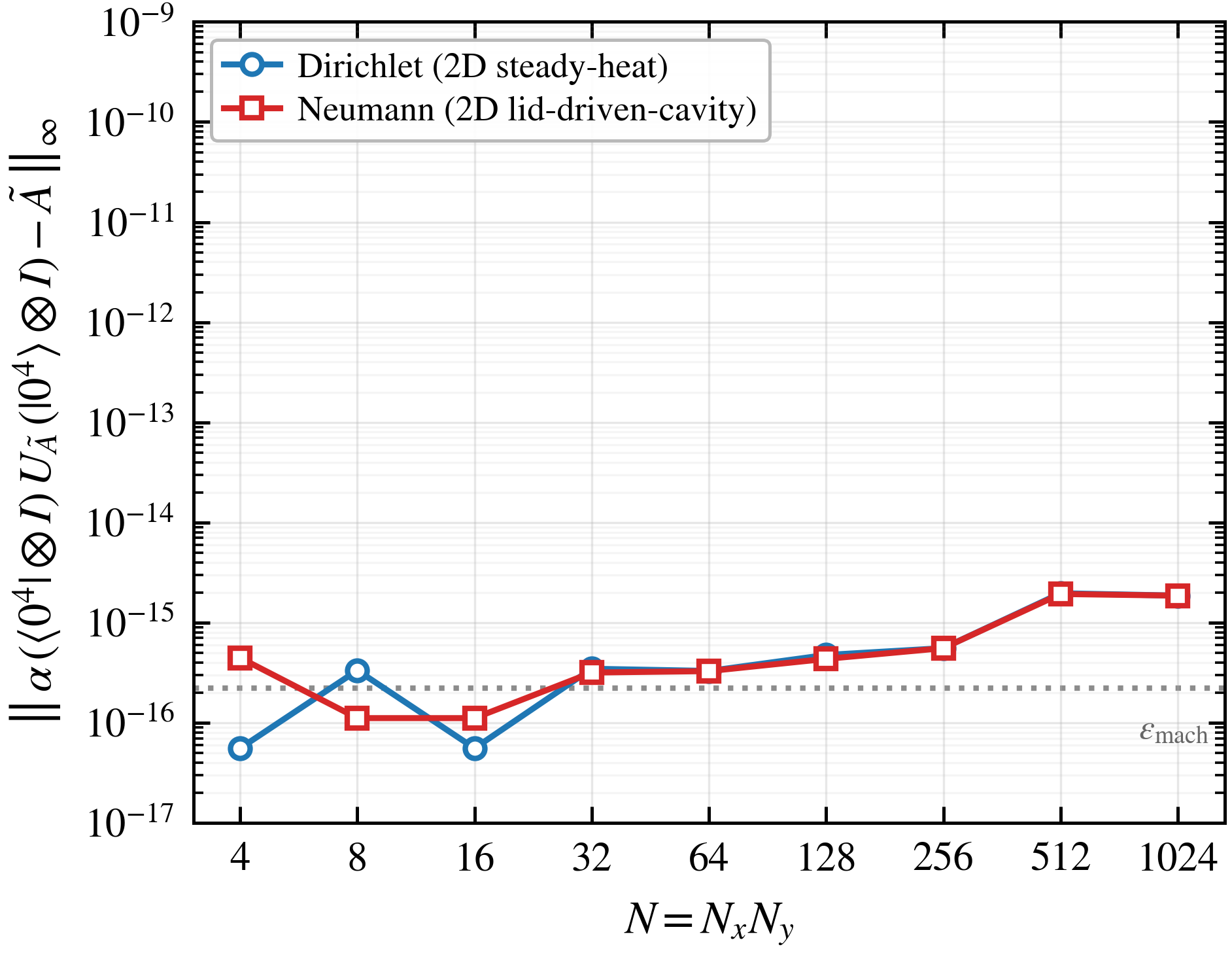}
    \caption{Infinity-norm deviation of the compiled circuit's top-left
  block from the target operator; the dotted line marks
  double-precision machine epsilon.}
  \label{fig:be_exactness}
  \end{subfigure}
  \hfill
  \begin{subfigure}[t]{0.48\textwidth}
    \centering
    \includegraphics[width=\textwidth]{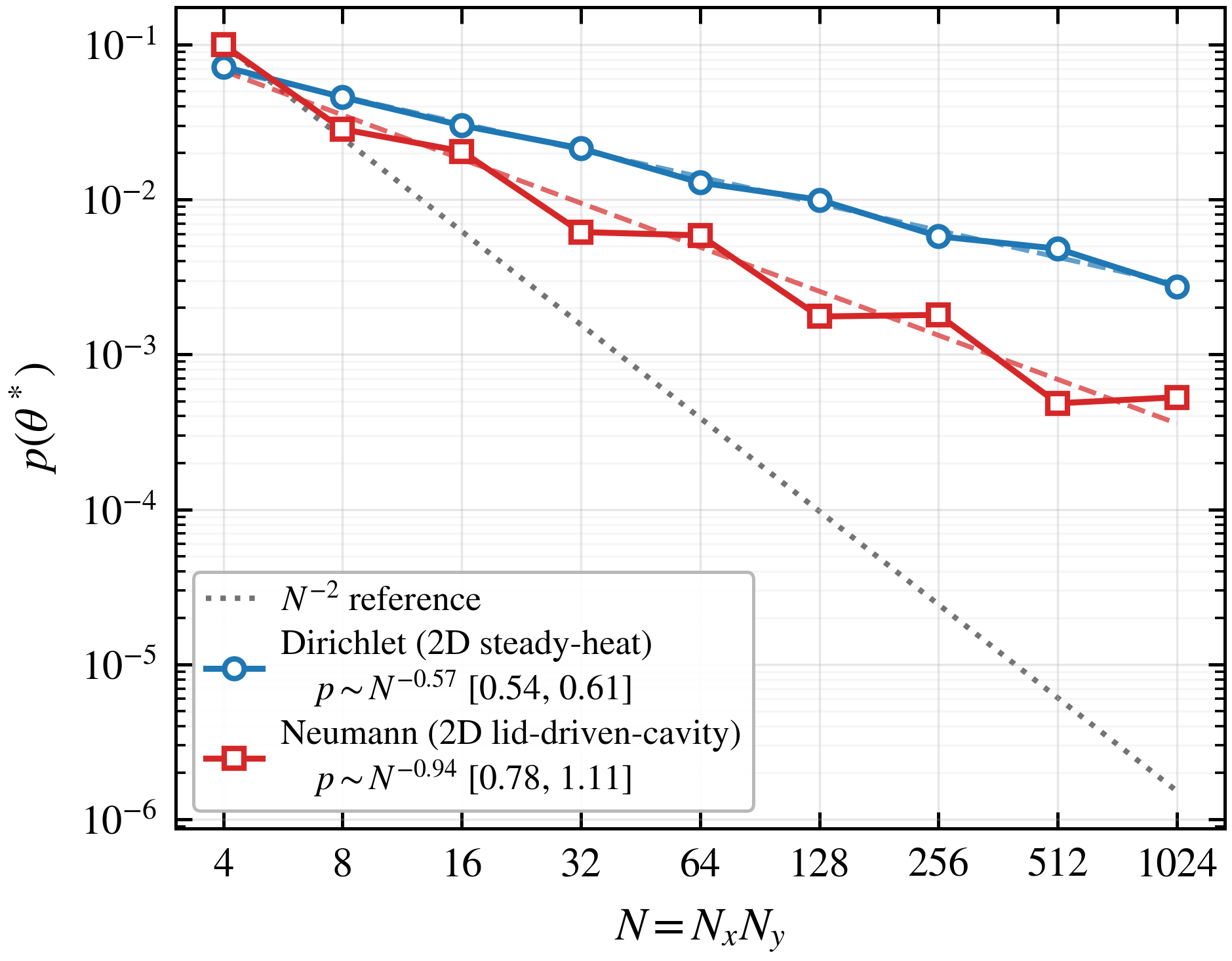}
    \caption{Post-selection success probability at the optimized
  parameters, with least-squares power-law fits $p\sim N^{-s}$ and $95\%$ confidence intervals.}
     \label{fig:be_success}
  \end{subfigure}

  \caption{Performance and numerical exactness of the block-encoding
  circuits used in this work.}
  \label{fig:be_results}
\end{figure*}

The construction of Section~\ref{subsec:block_encoding} is
an exact $(1,4,0)$-block encoding for both boundary conditions.
Figure~\ref{fig:be_exactness} verifies this directly by extracting the
top-left block of the assembled circuit unitary and comparing it against
the target operator.  The deviation
$\|(\langle 0^4|\otimes I)\,U_{\widetilde{A}}\,(|0^4\rangle\otimes I)
- \widetilde{A}\|_\infty$ remains at the level of double-precision
machine epsilon across the full range $N = 4$ to $N = 1024$, for the
Dirichlet operator of the heat conduction benchmark and the Neumann
operator of the lid-driven cavity alike.  

\subsubsection{Post-selection success probability}

Evaluating $C_G(\bm{\theta})$ through~\eqref{eq:cost_final} requires
post-selecting the ancilla register on $|0\rangle^{\otimes 4}$, and only
that fraction of shots contributes.  Since estimating $C_G$ to relative
precision $\varepsilon$ costs $\mathcal{O}(1/(p\varepsilon^2))$ shots,
the scaling of $p$ with system size determines whether the reduction to
a single circuit translates into a genuine saving.
Figure~\ref{fig:be_success} reports $p(\bm{\theta}^*)$ at the optimized
parameters.  The decay is far slower than the $N^{-2}$ behaviour that
the shrinking spectral gap of the scaled Laplacian might suggest:
fitting a power law gives $p \sim N^{-0.57}$ for the Dirichlet operator
and $p \sim N^{-0.94}$ for the Neumann operator.  The shot cost per
cost-function evaluation therefore grows as $\mathcal{O}(N^{0.57})$ and
$\mathcal{O}(N^{0.94})$ respectively, against $\Theta(N)$ circuits for
the Pauli-LCU formulation, each of which itself requires
$\mathcal{O}(1/\varepsilon^2)$ shots.

\subsubsection{Comparison with the Pauli-LCU formulation}

We compare against VQLS in the formulation
of~\cite{bravoprieto2023}, in which the Dirichlet Laplacian is expanded
in the Pauli basis and the cost function assembled from Hadamard tests.
The comparison uses the heat conduction operator; the Neumann case
behaves analogously and is omitted for brevity.  For the LCU
implementation we report, at each size, the resources of the most
expensive circuit in the set, which bounds the per-circuit requirement.

\begin{figure*}[t]
\centering
\begin{subfigure}[t]{0.32\textwidth}
  \centering
  \includegraphics[width=\textwidth]{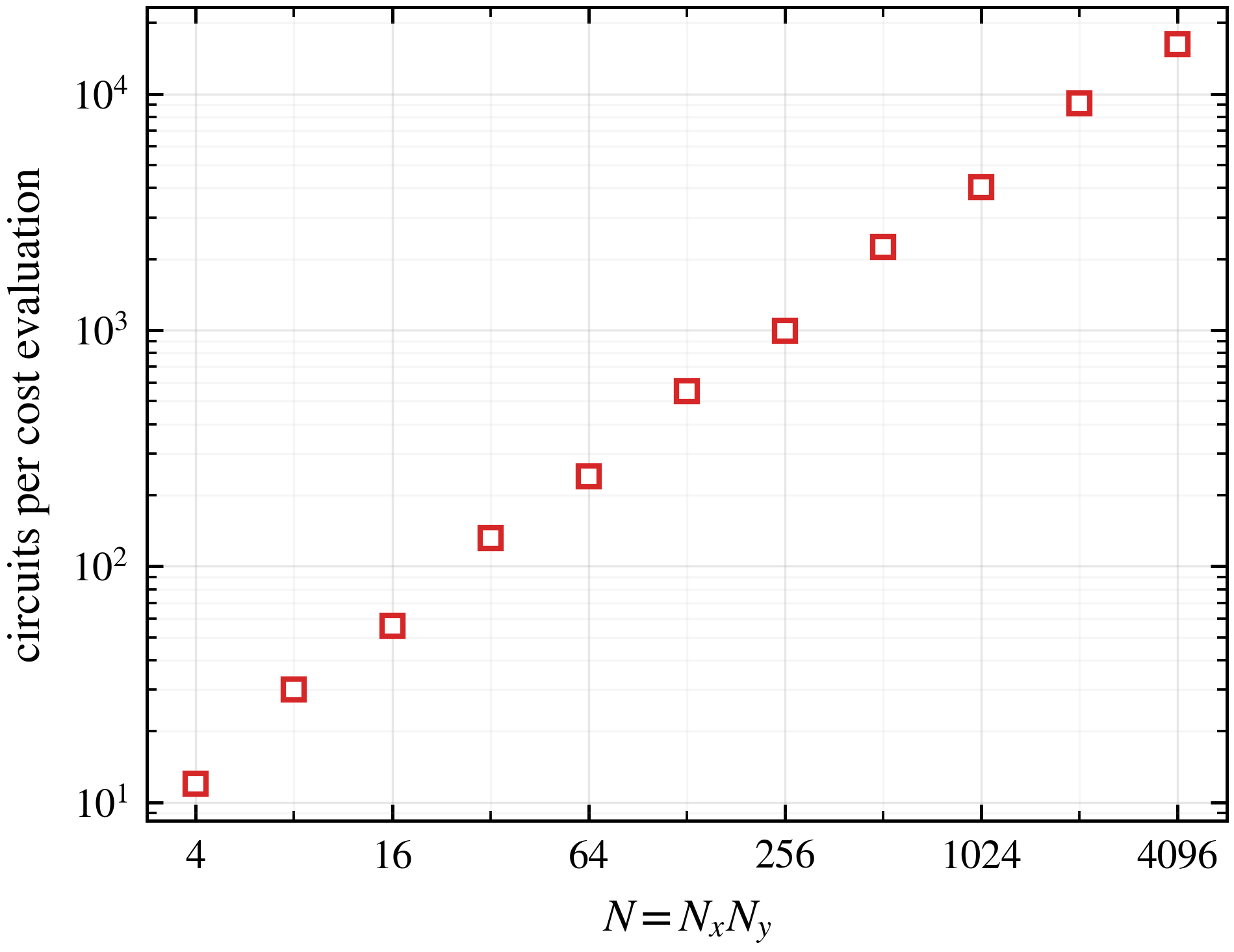}
  \caption{Circuits per cost evaluation, Pauli-LCU.}
  \label{fig:circuits_per_cost}
\end{subfigure}
\hfill
\begin{subfigure}[t]{0.32\textwidth}
  \centering
  \includegraphics[width=\textwidth]{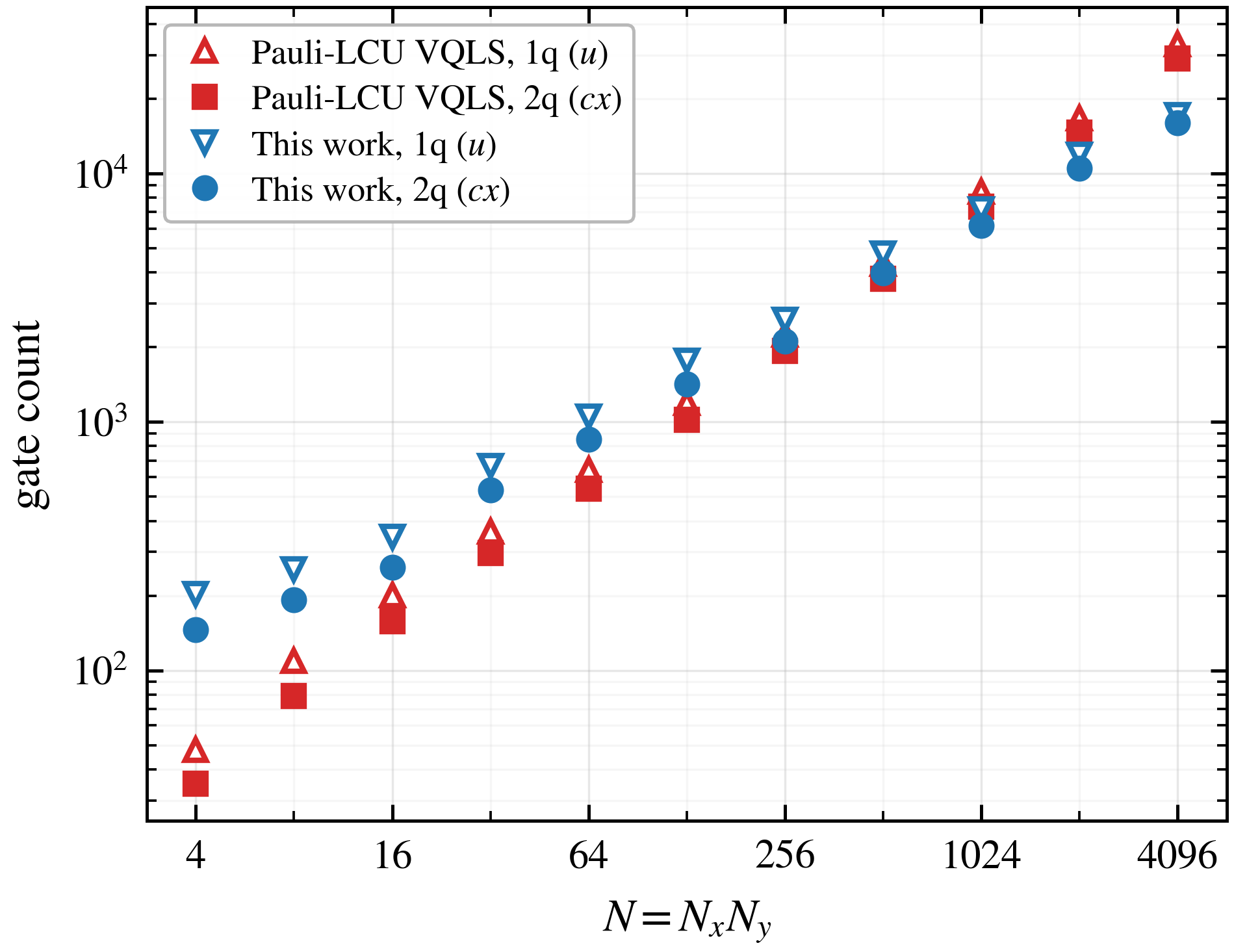}
  \caption{One- and two-qubit gate counts.}
  \label{fig:gate_count_comparison}
\end{subfigure}
\hfill
\begin{subfigure}[t]{0.32\textwidth}
  \centering
  \includegraphics[width=\textwidth]{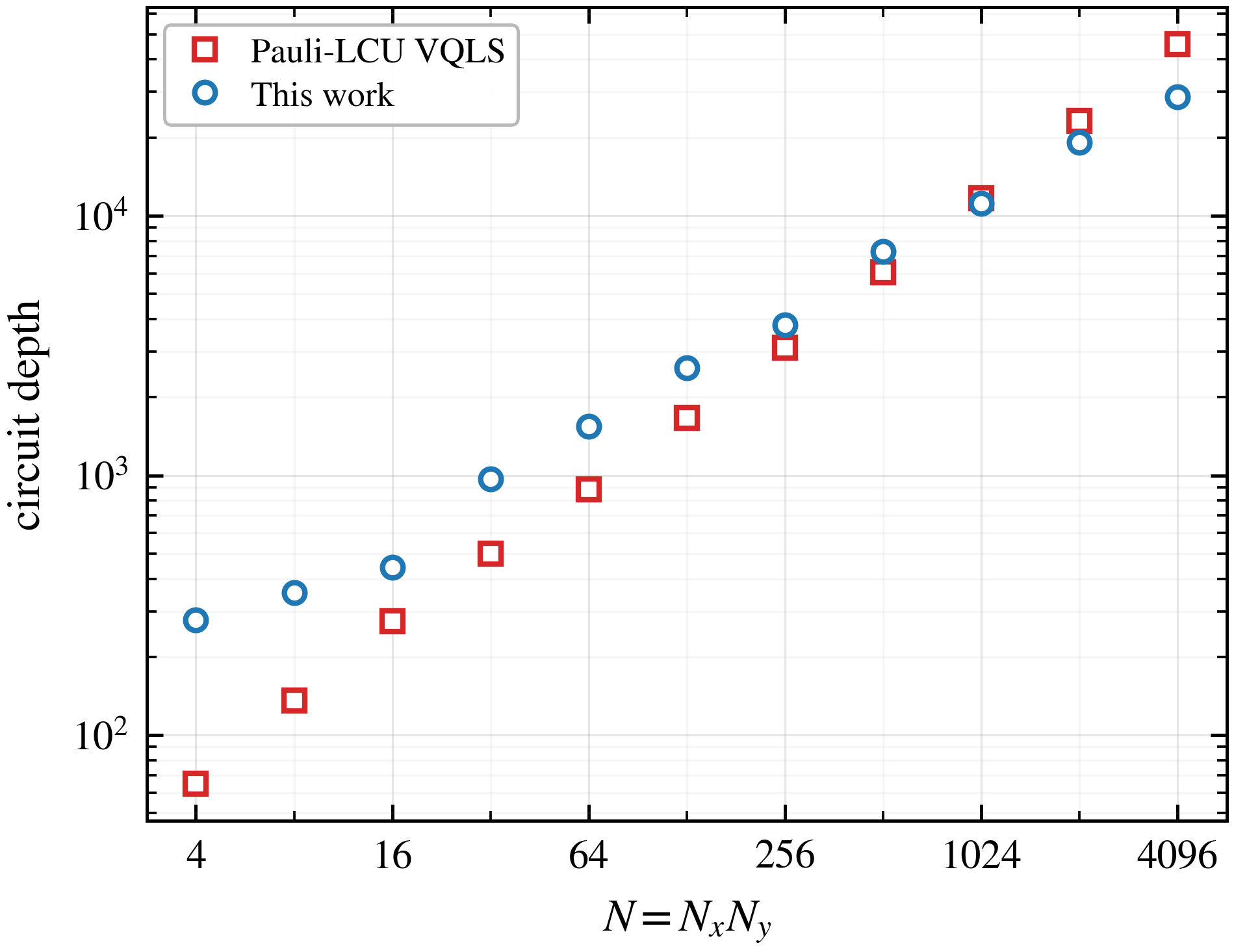}
  \caption{Circuit depth.}
  \label{fig:depth_comparison}
\end{subfigure}
\caption{Resource comparison between the Pauli-LCU VQLS and the
block-encoded formulation of this work, for the 2D Dirichlet Laplacian.
(a) The block-encoded formulation requires one circuit at every size
and is therefore not plotted.  (b), (c) Per-circuit counts after
transpilation to the $\{U, \mathrm{CX}\}$ basis; for the Pauli-LCU
formulation these are for the most expensive Hadamard-test circuit at
each grid, an empirical per-circuit upper bound.}
\label{fig:resource_comparison}
\end{figure*}

The principal difference is the number of circuits.  A Pauli
decomposition of the 2D Dirichlet Laplacian yields exactly
$L = N_x + N_y - 1$ terms, and the cost function requires $L(L+1)$
distinct Hadamard-test circuits per evaluation (Figure~\ref{fig:circuits_per_cost}). On a square grid this
is $\Theta(N)$, growing from $12$ circuits at $N = 4$ to $9120$ at
$N = 2048$, whereas the block-encoded formulation requires exactly one
circuit at every size.

Per-circuit resources tell a more nuanced story.  The block-encoded
circuit carries four ancillae against one for the LCU circuits, so its
width is larger by three qubits at every size.  In gate count and depth the LCU circuits are
individually cheaper at small and intermediate sizes, but the two
constructions scale differently and the ordering reverses beyond
$N = 512$ (Figures~\ref{fig:gate_count_comparison}
and~\ref{fig:depth_comparison}).  At $N = 1024$ the block-encoded
circuit is already the shallower of the two, and the margin widens with
system size: at $N = 4096$ it requires $16{,}034$ two-qubit gates
against $28{,}980$ for the heaviest LCU circuit, at a depth of
$28{,}741$ against $45{,}744$.

Taken together with the circuit count, the aggregate cost per
cost-function evaluation, defined as the product of the number of circuits and the gate count per-circuit, differs by orders of magnitude and the gap
widens monotonically with the size of the system. The block-encoded formulation is thus a cheaper construction on every axis except for a modest constant increase in width.

The Pauli decomposition is in one respect the most favourable LCU available: its per-circuit depth for the Hadamard test circuits used in VQLS is close to the minimum any linear-combination
formulation can achieve.  Its weakness is the term count, which forces $\Theta(N)$ circuit evaluations per cost function for the 2D Laplacian. More compact decompositions of the discrete Laplacian exist~\cite{hogancamp2026linearcombinationunitariesdecomposition}, reducing $L$ at the cost of unitaries that are
individually more expensive.  Since the block-encoded circuit is already shallower than the Pauli-LCU circuit beyond $N = 512$, and any such decomposition has a per-circuit depth at least that of the Pauli case, the crossover against a compact LCU would be expected at a smaller system size.  

\subsection{Solution Accuracy}
\label{subsec:accuracy}

Three benchmark problems are considered.  The first two are instances of
the Dirichlet problem~\eqref{eq:poisson} and differ only in the source
term; the third places the solver inside a time-advancing flow
simulation under Neumann boundary conditions.  Accuracy is reported as
the infidelity $1 - F$ between the recovered and reference solution
vectors, where $F = |\langle x^{\mathrm{SOR}} | x(\bm{\theta}^*)
\rangle|^2$. All runs use CMA-ES optimizer with $\sigma_0 = 0.3$ and a population of $15$ per
generation, terminating at $6 \times 10^{4}$ generations or on a cost
tolerance of $10^{-12}$, whichever is reached first.  The ansatz layers
is $\ell = 3$ for $N \leq 256$ and $\ell = 5$ at $N = 512$ and
$N = 1024$.

\subsubsection{Sinusoidal source}
\label{subsubsec:sine_results}

The first benchmark takes the smooth source term
\begin{equation}
  f(x,y) = -\sin(\pi x)\sin(\pi y),
  \label{eq:sine_source}
\end{equation}
with homogeneous Dirichlet conditions on all four walls.  The forcing is
smooth and free of singularities, and the problem admits the analytical
solution $u(x,y) = \sin(\pi x)\sin(\pi y)/(2\pi^2)$.

\begin{figure*}[t]
\centering
\begin{subfigure}[t]{0.32\textwidth}
    \centering
    \includegraphics[width=\textwidth]{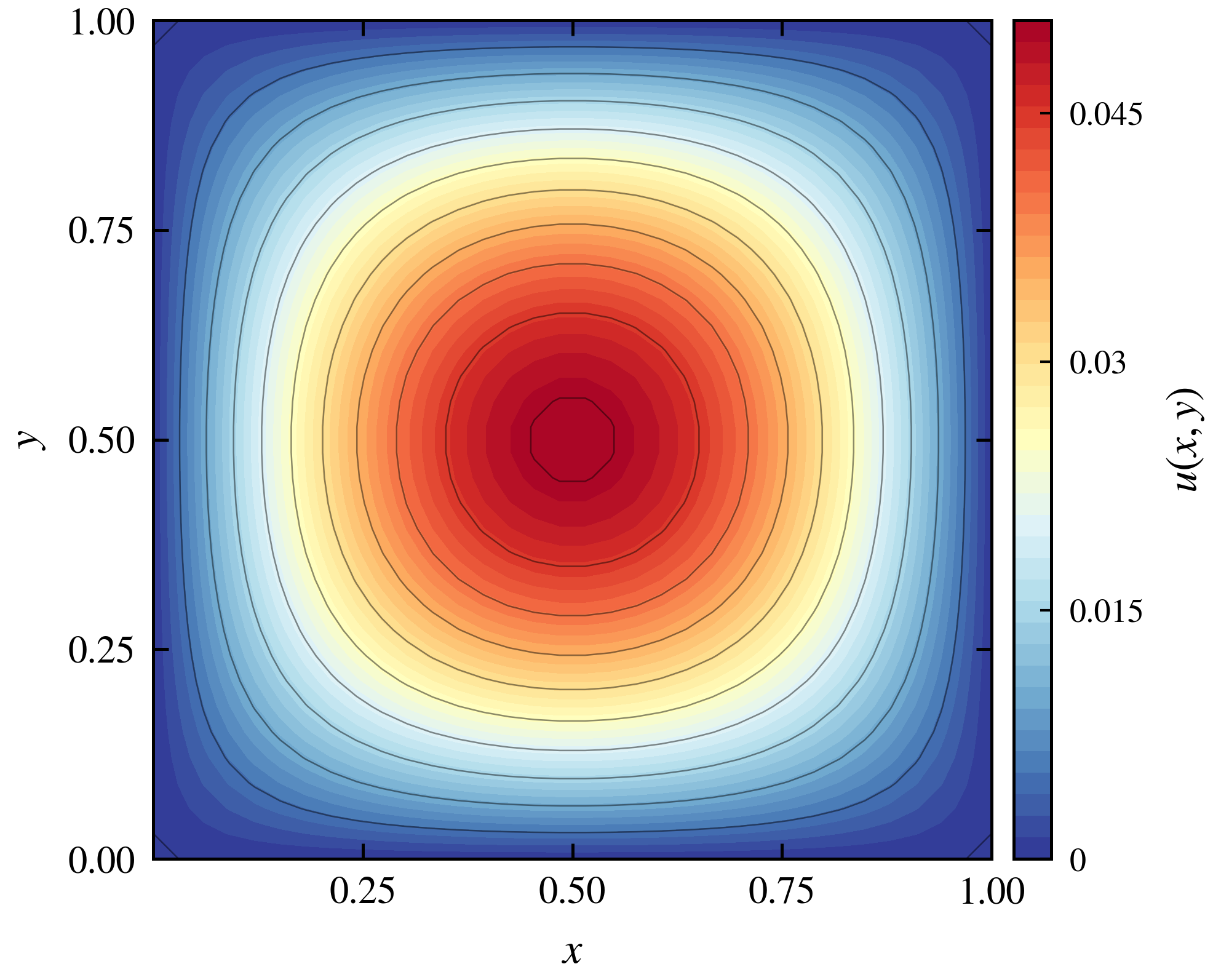}
  \caption{Classical (SOR) solution.}
\end{subfigure}
\hfill
\begin{subfigure}[t]{0.32\textwidth}
    \centering
    \includegraphics[width=\textwidth]{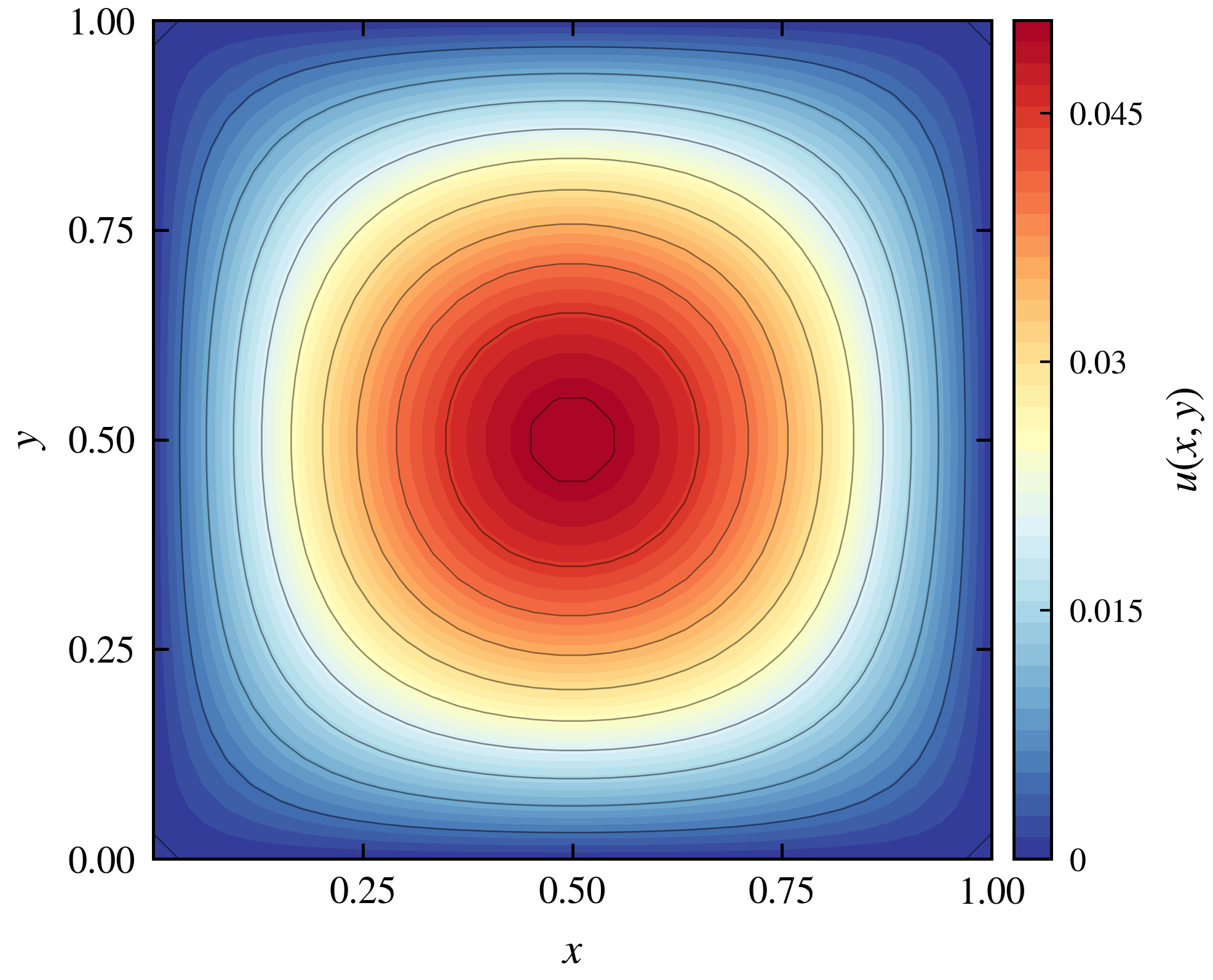}
  \caption{VQLS solution, median seed.}
\end{subfigure}
\hfill
\begin{subfigure}[t]{0.32\textwidth}
    \centering
    \includegraphics[width=\textwidth]{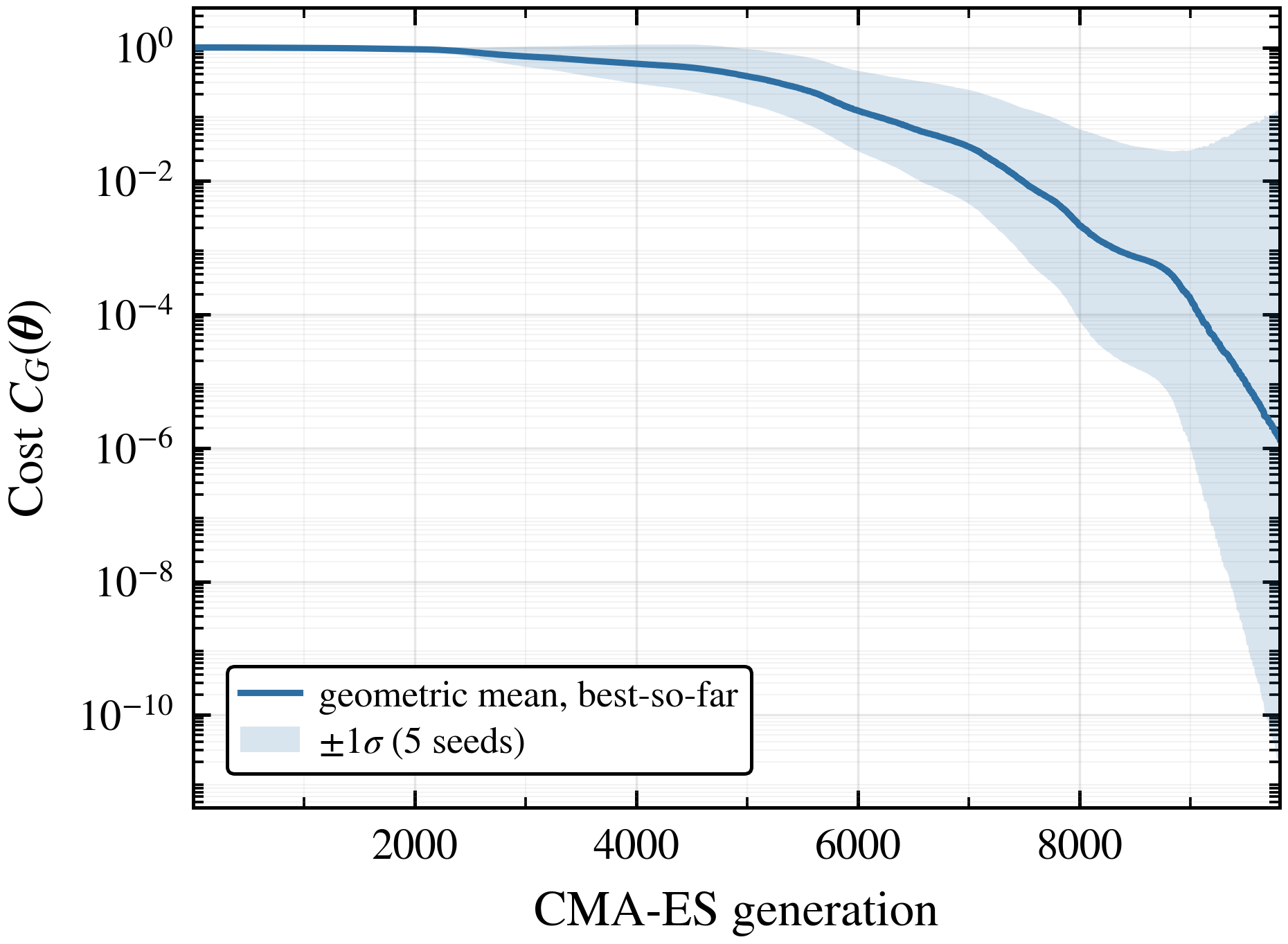}
  \caption{Cost history over five seeds.}
\end{subfigure}
\caption{Sinusoidal source benchmark on the $32 \times 32$ grid
($N = 1024$, $n = 10$ system qubits, $14$ total), CMA-ES with
$\sigma_0 = 0.3$ and $\ell = 5$ ansatz layers.  Panels (a) and (b) share
a colour scale and show the seed attaining the median infidelity of the
five runs, $1 - F = 8.9 \times 10^{-16}$.  In (c) the solid line is the
geometric mean of the best-so-far cost across seeds and the band is
$\pm 1$ standard deviation in $\log_{10}$; the median seed reaches
$C_G = 4.7 \times 10^{-15}$.}
\label{fig:sine_1024}
\end{figure*}

Figure~\ref{fig:sine_1024} shows the classical and VQLS solution fields
on the $32 \times 32$ grid ($N = 1024$, $n = 10$ system qubits, $14$
total), together with the cost-function history averaged over the five
seeds.  The VQLS solution reproduces the smooth bell-shaped
distribution characteristic of the sinusoidal source, peaking at the
domain centre and decaying to zero at the walls in accordance with the
homogeneous Dirichlet conditions. Agreement is exact to arithmetic precision: the median seed attains
$1 - F = 8.9 \times 10^{-16}$ at a final cost of
$C_G = 4.7 \times 10^{-15}$, and all five seeds lie at the same level.

\subsubsection{Gaussian source}
\label{subsubsec:heat_results}

The second benchmark is the steady-state heat conduction equation
$-k\nabla^2 T = Q(x,y)$, an instance of~\eqref{eq:poisson} with
$u \equiv T$ and a spatially localized source,
\begin{equation}
  Q(x,y)
  = Q_0 \exp\!\left(
      -\frac{(x-x_0)^2 + (y-y_0)^2}{2\sigma^2}
    \right),
  \label{eq:gaussian_source}
\end{equation}
with $Q_0 = 5$, $(x_0, y_0) = (0.5, 0.3)$ and $\sigma = 0.1$.  The
boundary conditions are Dirichlet throughout: the bottom wall is held at
$T_\mathrm{hot} = 1$, the top wall at $T_\mathrm{cold} = 0$, and the
side walls at the mean of the two.  The configuration models a
heat-generating component embedded in a conducting plate held between a
hot base and a cold lid, and unlike the sinusoidal case it presents the
solver with a sharp interior gradient superposed on a smooth background
stratification.

\begin{figure*}[t]
\centering
\begin{subfigure}[t]{0.32\textwidth}
    \centering
    \includegraphics[width=\textwidth]{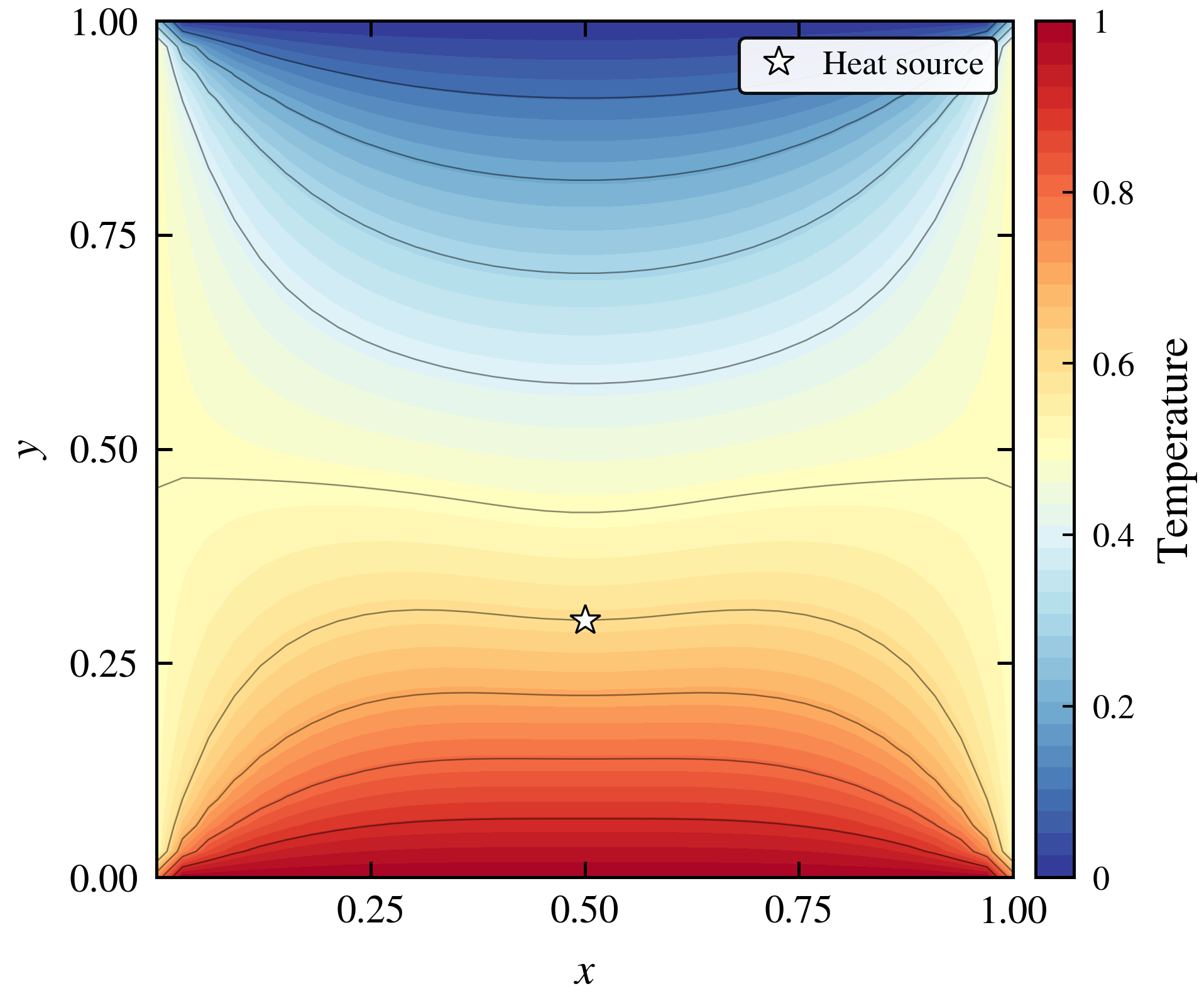}
    \caption{Classical (SOR) solution.}
  \end{subfigure}
  \hfill
  \begin{subfigure}[t]{0.32\textwidth}
    \centering
    \includegraphics[width=\textwidth]{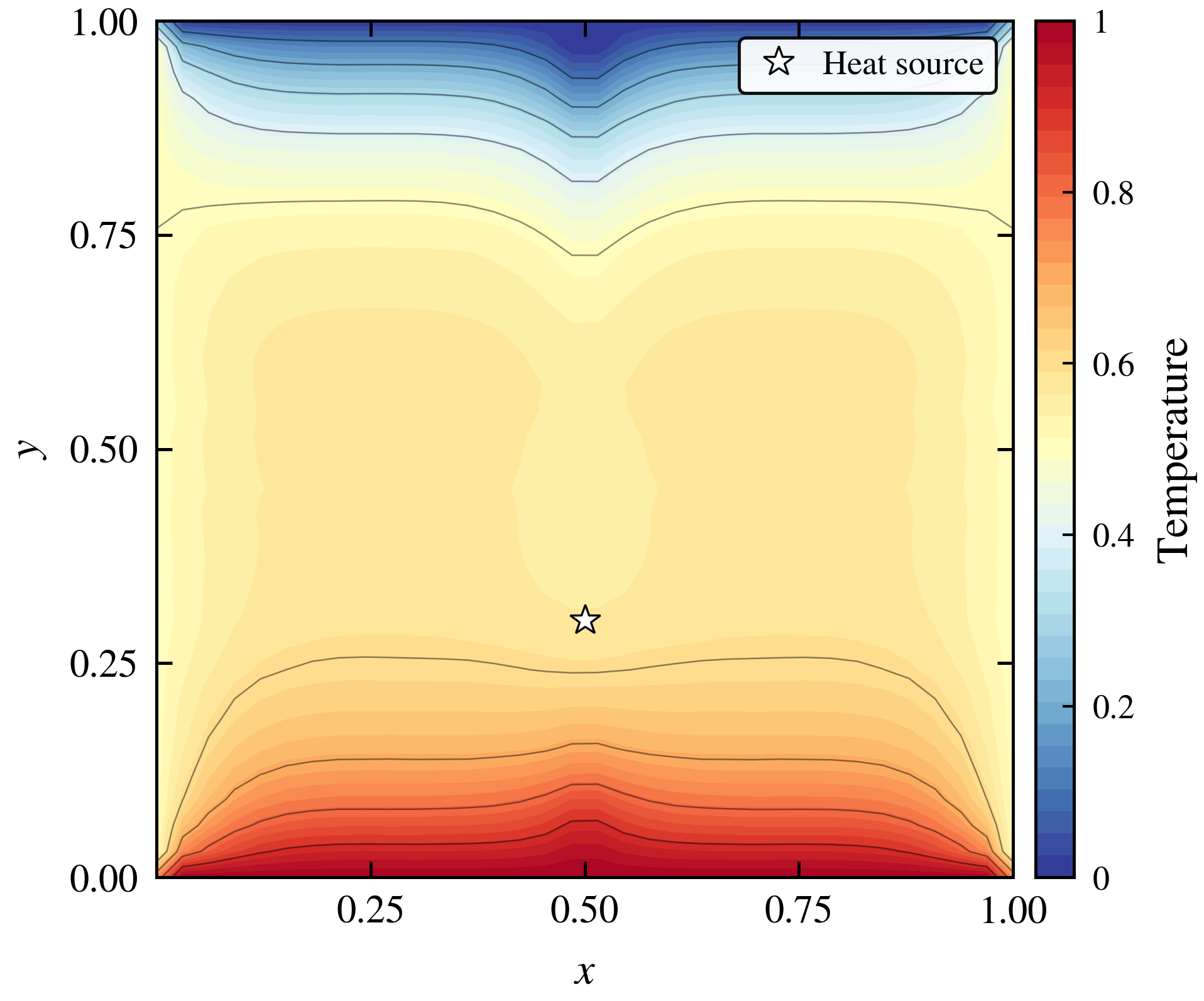}
    \caption{Quantum (VQLS) solution.}
  \end{subfigure}
  \hfill
 \begin{subfigure}[t]{0.32\textwidth}
    \centering
    \includegraphics[width=\textwidth]{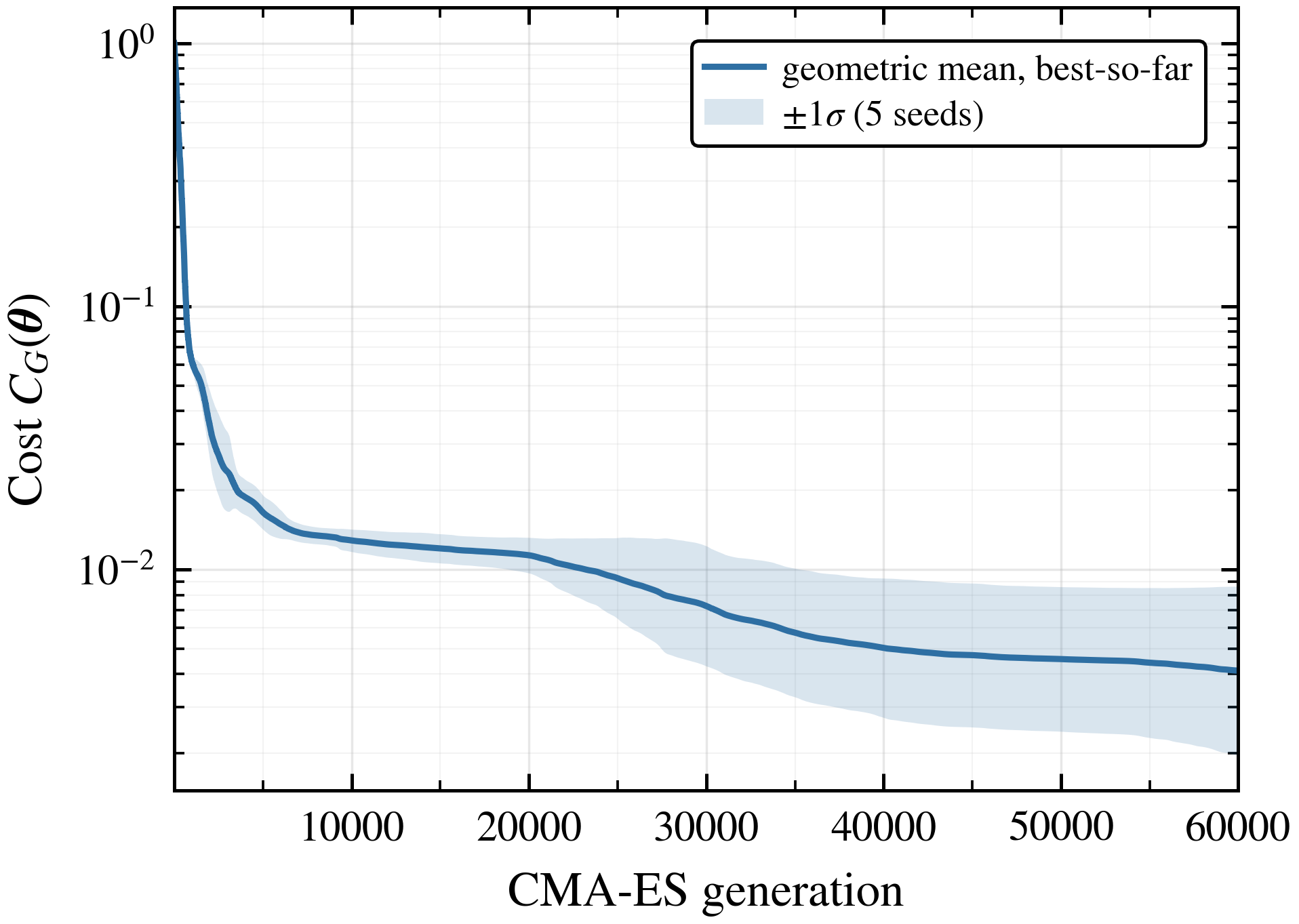}
    \caption{Cost history over five seeds.}
  \end{subfigure}
  \caption{Heat conduction benchmark with Gaussian source on the $32 \times 32$ grid ($N = 1024$, $n = 10$ system qubits, $14$ total), CMA-ES with $\sigma_0 = 0.3$ and $\ell = 5$ ansatz layers.  The star marks the source at $(x_0, y_0) = (0.5, 0.3)$.  Panels (a) and (b) share a colour scale and show the seed attaining the median infidelity of the five runs, $1 - F = 4.4 \times 10^{-2}$.  In (c) the solid line is the geometric mean of the best-so-far cost across seeds and the band is $\pm 1$ standard deviation in $\log_{10}$; the median seed reaches $C_G = 4.2 \times 10^{-3}$.}
  \label{fig:heat_1024}
\end{figure*}

Figure~\ref{fig:heat_1024} shows the corresponding fields and cost
history on the $32 \times 32$ grid.
The VQLS field reproduces the vertical thermal stratification set by the
hot and cold walls, but the isotherms are flattened relative to the
classical reference.  The median seed attains $1 - F = 4.4 \times 10^{-2}$ at a final
cost of $C_G = 4.2 \times 10^{-3}$.

Table~\ref{tab:accuracy} summarises the accuracy attained on both
Dirichlet benchmarks across the full range of system sizes.  The two
behave quite differently.  The sinusoidal source is recovered to
arithmetic precision at every size, which is expected once one notes
that the discretized forcing is an exact eigenvector of the discrete
Dirichlet Laplacian: the solution state coincides with $|b\rangle$,
so the ansatz need only prepare a state of Schmidt rank one across the
$x$--$y$ register cut.  This case therefore verifies the pipeline
end to end. Fo the Gaussian source, the
solution has an overlap of $\approx 0.78$ with $|b\rangle$ at $N=1024$.  Its infidelity remains at arithmetic precision through
$N = 32$, and rises for grid sizes beyond this.  This can be attributed to the optimization which is discussed in Section~\ref{subsec:optimizer_results}.

\begin{table}[!t]
  \centering
  \caption{VQLS solution infidelity $1 - F$ across system sizes for both
    Dirichlet benchmarks, where
    $F = |\langle x^{\mathrm{SOR}} | x(\bm{\theta}^*)\rangle|^2$.
    Entries are the geometric mean over five independent seeds, for
    CMA-ES with $\sigma_0 = 0.3$ at ansatz depth $\ell$.
    $n$ denotes system qubits; total circuit width is $n + 4$.}
  \label{tab:accuracy}
  \begin{tabular}{cccccc}
    \toprule
    \multirow{2}{*}{$N_x \times N_y$}
    & \multirow{2}{*}{$N$}
    & \multirow{2}{*}{$n$}
    & \multirow{2}{*}{$\ell$}
    & \multicolumn{2}{c}{$1 - F$} \\
    \cmidrule(lr){5-6}
    & & & & Sine & Heat \\
    \midrule
    $2  \times 2$  &    4 &  2 & 3 & $3.1 \times 10^{-16}$ & $3.0 \times 10^{-16}$ \\
    $2  \times 4$  &    8 &  3 & 3 & $2.1 \times 10^{-16}$ & $1.1 \times 10^{-15}$ \\
    $4  \times 4$  &   16 &  4 & 3 & $3.8 \times 10^{-16}$ & $1.8 \times 10^{-15}$ \\
    $4  \times 8$  &   32 &  5 & 3 & $3.2 \times 10^{-16}$ & $1.5 \times 10^{-15}$ \\
    $8  \times 8$  &   64 &  6 & 3 & $3.4 \times 10^{-16}$ & $3.2 \times 10^{-5}$  \\
    $8  \times 16$ &  128 &  7 & 3 & $2.5 \times 10^{-16}$ & $1.4 \times 10^{-3}$  \\
    $16 \times 16$ &  256 &  8 & 3 & $1.8 \times 10^{-16}$ & $3.8 \times 10^{-2}$  \\
    $16 \times 32$ &  512 &  9 & 5 & $1.2 \times 10^{-16}$ & $3.8 \times 10^{-2}$  \\
    $32 \times 32$ & 1024 & 10 & 5 & $2.0 \times 10^{-15}$ & $5.1 \times 10^{-2}$  \\
    \bottomrule
  \end{tabular}
\end{table}

\subsubsection{Lid-driven cavity}
\label{subsubsec:ldc_results}

Here we empirically examine, whether the accuracy attainable by the VQLS solver suffices to sustain a simulation in which the system
is solved repeatedly, with a right-hand side that evolves as the
solution does.

We take the two-dimensional lid-driven cavity, the standard validation
case for incompressible flow solvers.  Fluid occupies a square domain
with no-slip conditions on all four walls, the upper wall translating at
constant velocity $U_\mathrm{lid}$ and driving a recirculating vortex in
the interior.  The flow is advanced by the projection method
of Section~\ref{subsec:poisson}: at each time step an intermediate
velocity field is predicted, the Neumann pressure-Poisson
system~\eqref{eq:kronecker_neumann} is solved to obtain the pressure
correction, and the velocity is projected onto the divergence-free
subspace~\cite{gresho1987}. The VQLS solver of Section~\ref{subsec:vqls_theory} is
invoked once per time step in place of the classical pressure solve,
using the Neumann block encoding of
Section~\ref{subsubsec:be_neumann}. Each step is warm started from the converged parameters of the previous
step, so only the first solve begins from a random initialization. Because the recovered pressure
enters the velocity correction directly, any error committed at one step
is carried into the initial condition of the next. 

The simulation is run on a $16 \times 16$ grid at $Re = 100$ with
$\Delta t = 0.05$, giving a Neumann system of $N = 256$ unknowns
($n = 8$ system qubits).  Because every time step
requires a complete variational optimization, we advance ten steps to
$t = 0.5$ and compare against the classical solver run.  At this time the
viscous diffusion length is $\sqrt{\nu t} \approx 0.07$, roughly one
cell width, so the flow is still developing and far away from the steady-state.

\begin{figure*}[t]
  \centering
  \begin{subfigure}[t]{0.48\textwidth}
    \centering
    \includegraphics[width=\textwidth]{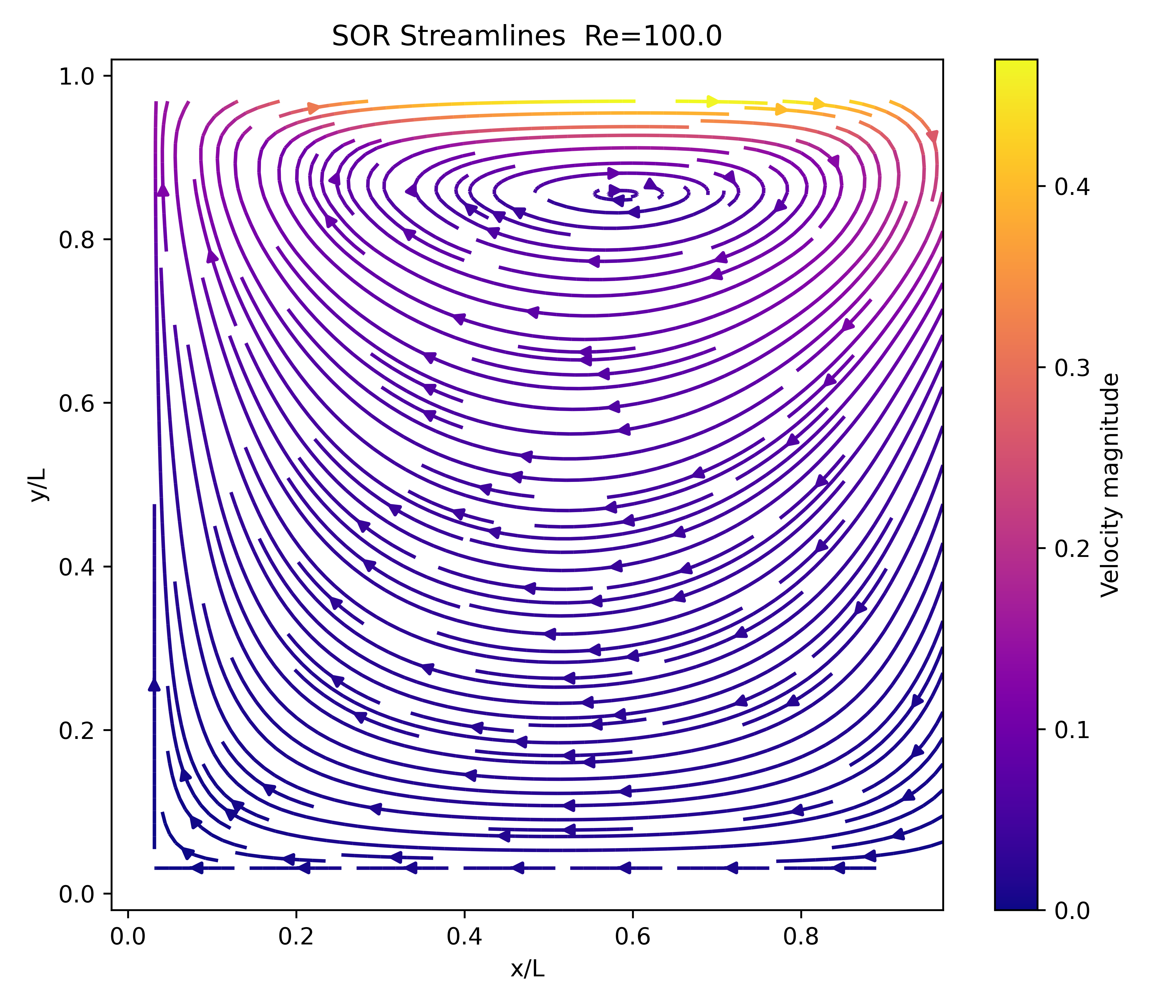}
    \caption{Classical (SOR) streamlines.}
    \label{fig:ldc_sor}
  \end{subfigure}
  \hfill
  \begin{subfigure}[t]{0.48\textwidth}
    \centering
    \includegraphics[width=\textwidth]{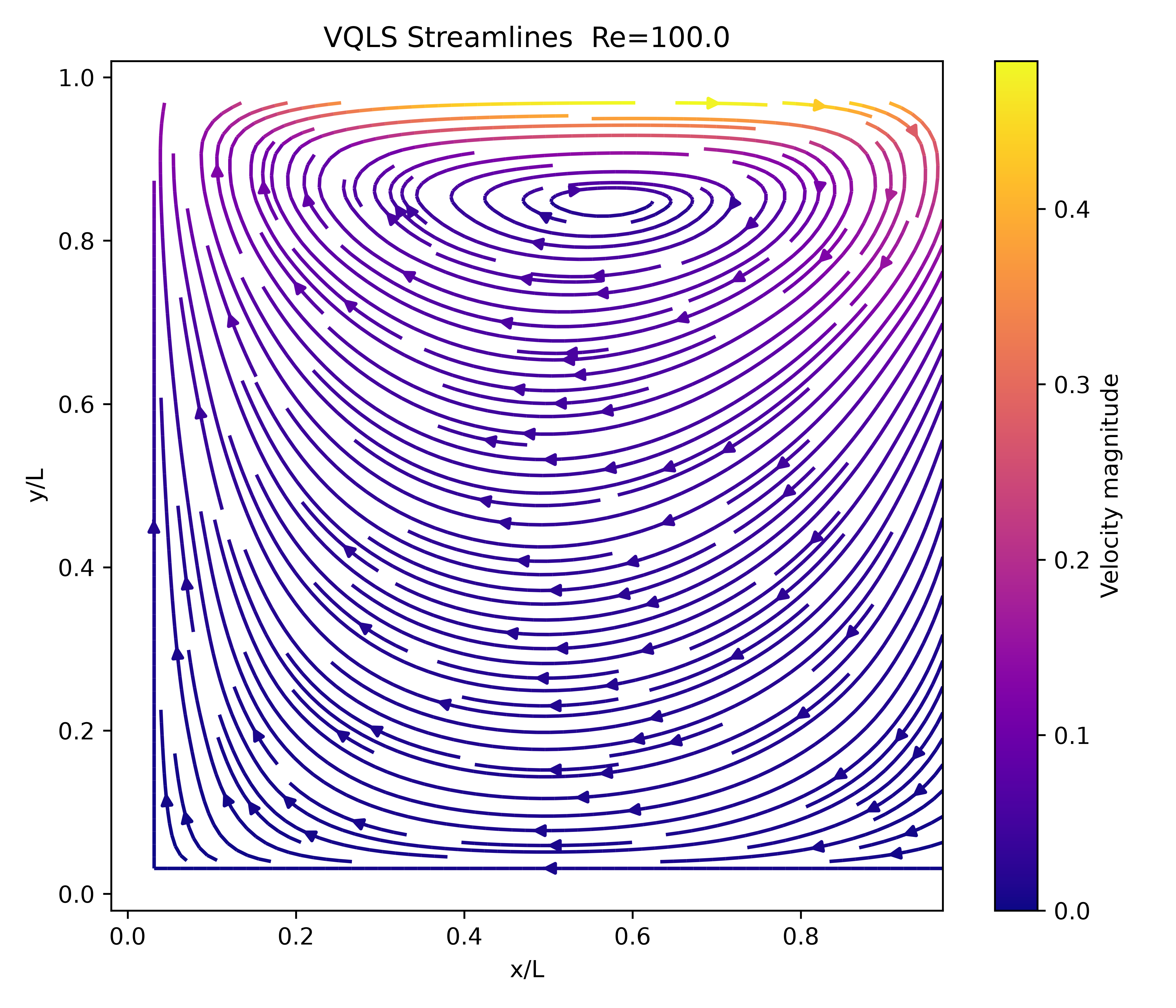}
    \caption{VQLS streamlines.}
    \label{fig:ldc_vqls}
  \end{subfigure}

  \vspace{0.6em}

  \begin{subfigure}[t]{0.48\textwidth}
    \centering
    \includegraphics[width=\textwidth]{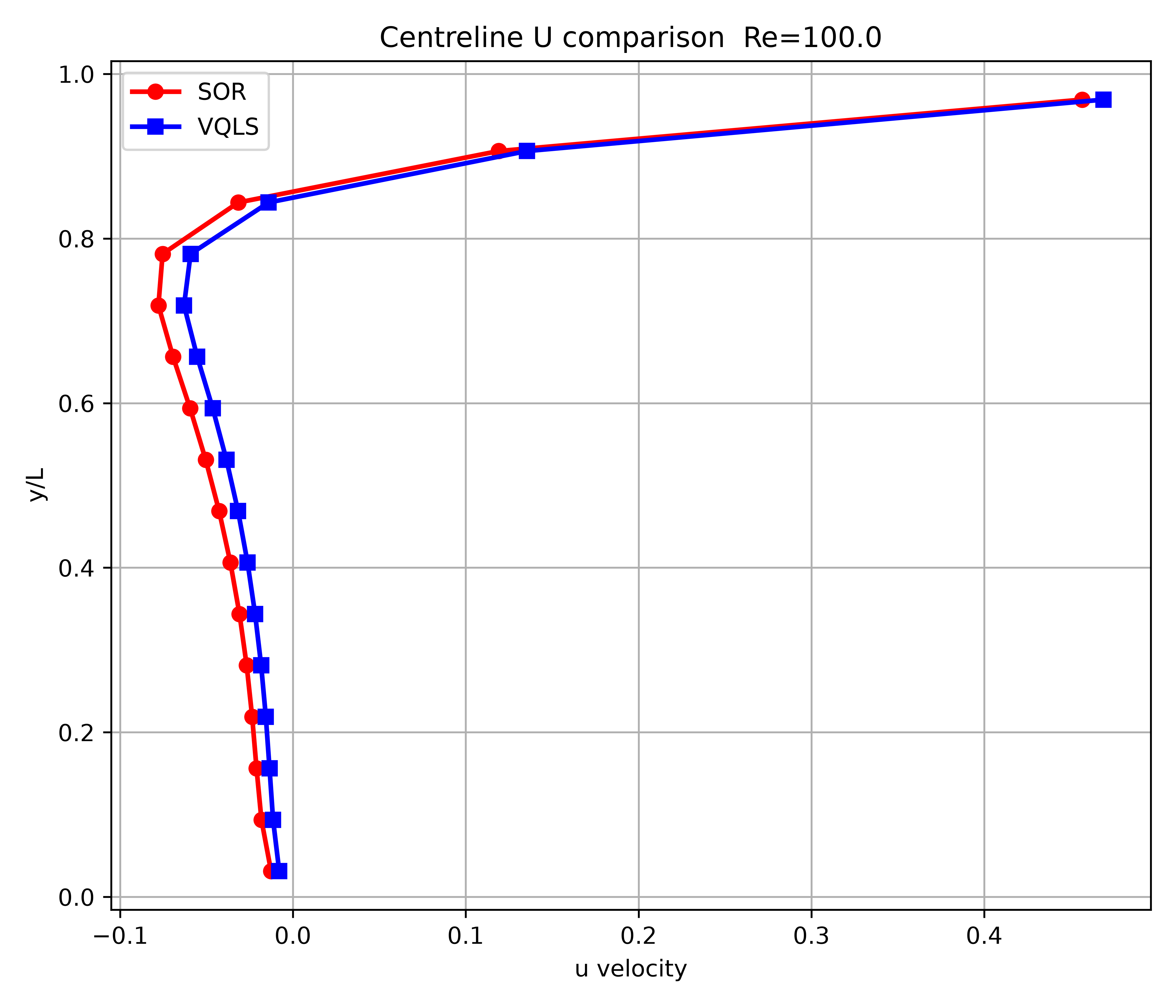}
    \caption{$u$ along the vertical centreline.}
    \label{fig:ldc_u}
  \end{subfigure}
  \hfill
  \begin{subfigure}[t]{0.48\textwidth}
    \centering
    \includegraphics[width=\textwidth]{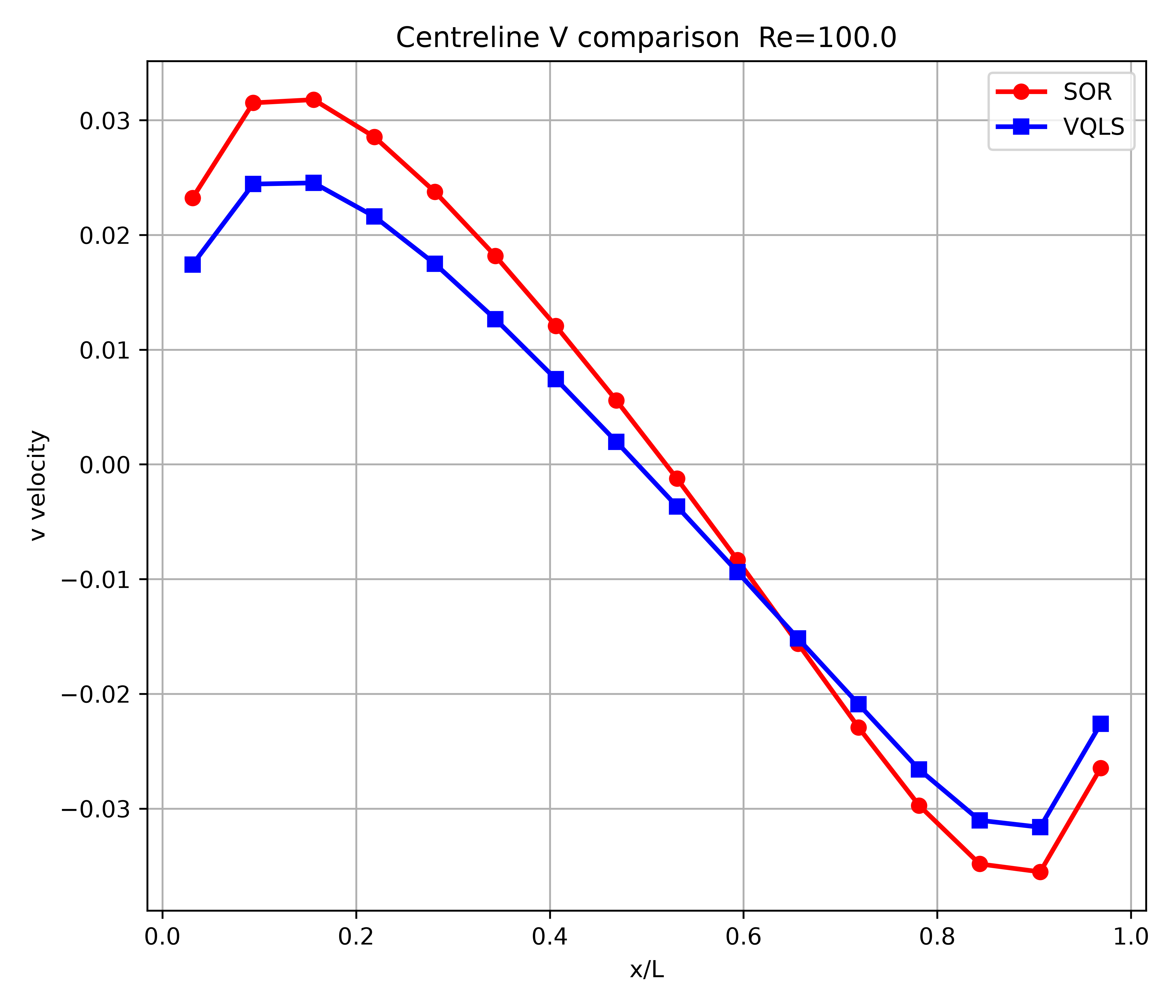}
    \caption{$v$ along the horizontal centreline.}
    \label{fig:ldc_v}
  \end{subfigure}

  \caption{Lid-driven cavity at $Re = 100$ on a $16 \times 16$ grid,
    after ten steps of $\Delta t = 0.05$ ($t = 0.5$).  The two solvers
    differ only in the pressure-Poisson solve; the discretization,
    initial condition and time step are identical.  The flow is still
    developing at this time.}
  \label{fig:ldc}
\end{figure*}

Figure~\ref{fig:ldc} shows the outcome.  The streamline patterns are
closely similar: both solvers place the primary vortex near
$(x/L, y/L) \approx (0.6, 0.85)$ and reproduce the same recirculation
structure.
The centreline profiles quantify the difference. The discrepancy is systematic and the quantum pressure solve slightly damps the
velocity magnitude while preserving the structure of the field.

Thus we observe that the two trajectories remain together over ten time steps.  Each step feeds its pressure directly into the
velocity correction, so an error committed at one step enters the
initial condition of the next; a solver whose output were merely
qualitatively correct would be expected to drift.

\subsection{Barren Plateaus and the Role of the Optimizer}
\label{subsec:optimizer_results}

The results of the 2D heat conduction problem degrade with system size as shown in Table~\ref{tab:accuracy}.  The limitation lies in the optimization, and we examine it here in two steps: first establishing that the cost landscape
exhibits a barren plateau in the sense of~\eqref{eq:barren}, then asking
what distinguishes an optimizer that continues to make progress from one
that does not.

\subsubsection{Gradient variance under uniform sampling}

Figure~\ref{fig:bp_variance_uniform} reports $\mathrm{Var}_{\bm{\theta}}[\partial_\mu C_G]$, averaged over parameter
indices $\mu$, for $\bm{\theta}$ drawn uniformly from $[0,2\pi)^d$, at
ansatz depths $\ell = 2$ to $5$ and system sizes $n = 2$ to $10$.  The
decay is exponential in $n$ at every depth,
which is the defining signature of~\eqref{eq:barren}.  Fitting
$\mathrm{Var} \propto b^{-n}$ gives $b = 3.90$, $3.78$, $3.77$ and
$3.95$ for $\ell = 2$ to $5$: statistically indistinguishable across
depths. This is consistent with the plateau being a
property of the global cost function~\eqref{eq:cost} rather than of
ansatz expressivity, the regime identified in~\cite{cerezo2021cost}.

\subsubsection{Gradient variance along the optimizer trajectory}

\begin{figure*}[t]
  \centering
  \begin{subfigure}[t]{0.48\textwidth}
    \centering
    \includegraphics[width=\textwidth]{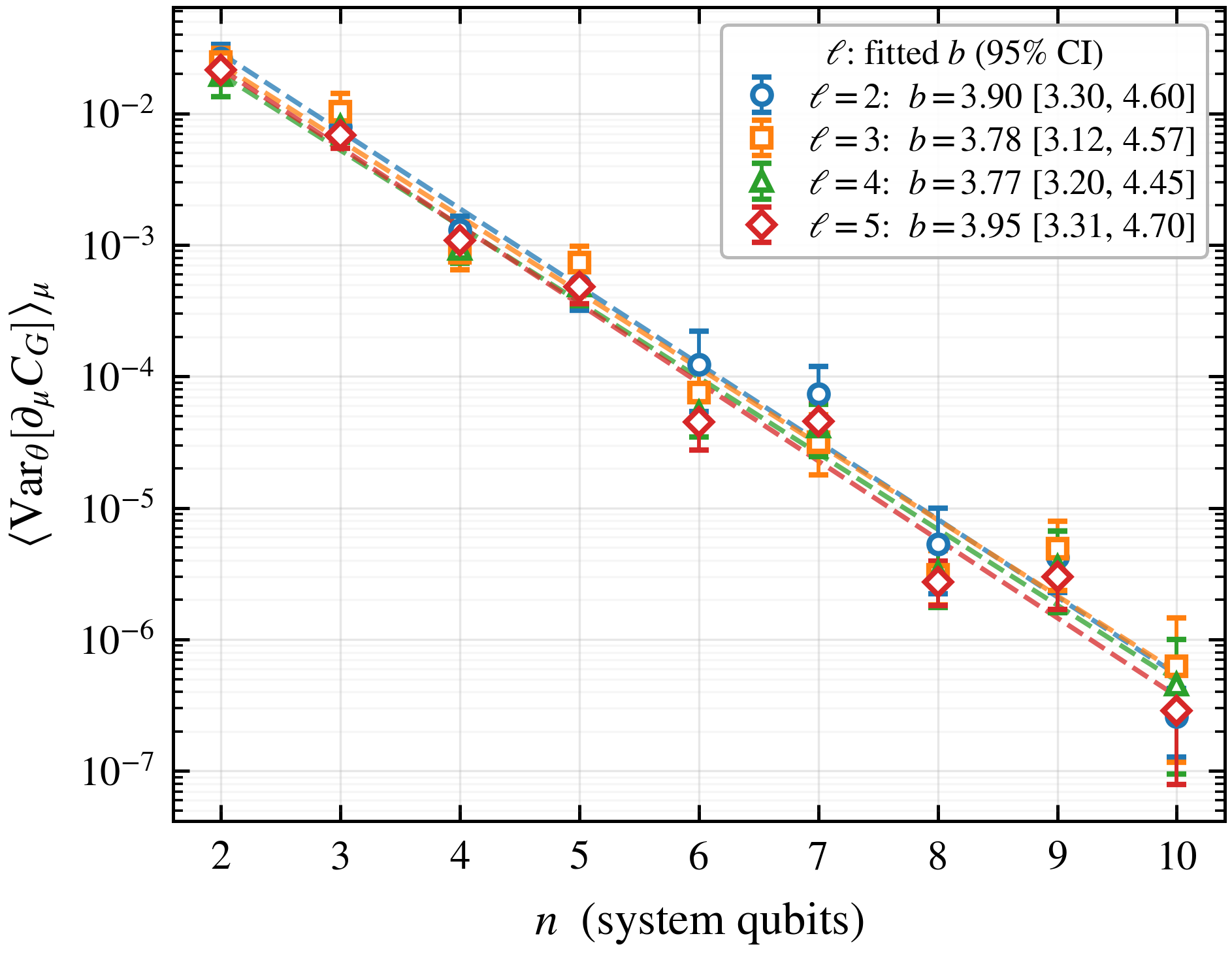}
    \caption{Uniform sampling, $\bm{\theta} \sim U[0,2\pi)^d$.
      Dashed lines are fits of $\mathrm{Var} \propto b^{-n}$.}
    \label{fig:bp_variance_uniform}
  \end{subfigure}
  \hfill
  \begin{subfigure}[t]{0.48\textwidth}
    \centering
    \includegraphics[width=\textwidth]{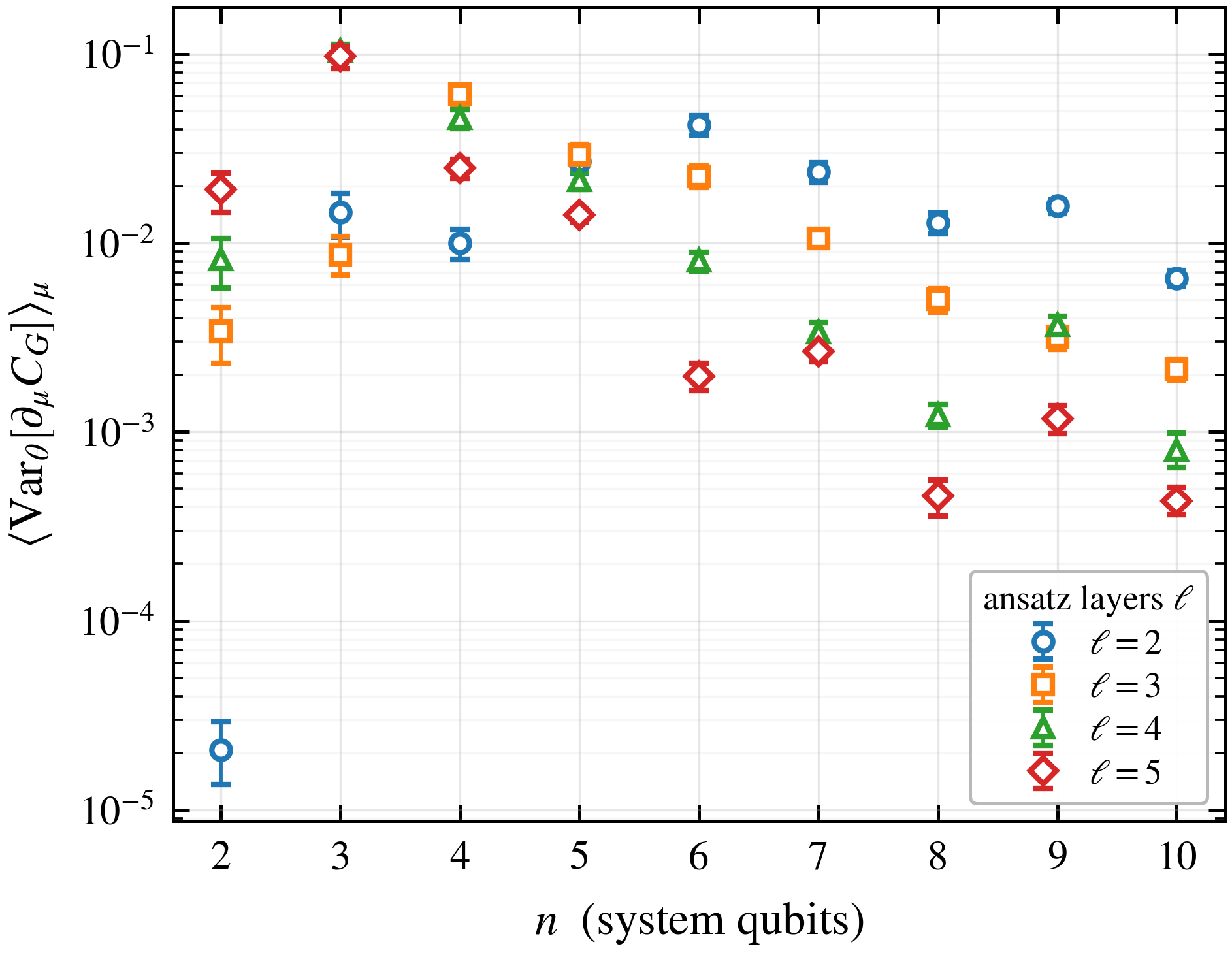}
    \caption{Sampling along the CMA-ES trajectory.  No fit is shown;
      see text.}
    \label{fig:bp_variance_trajectory}
  \end{subfigure}
  \caption{Variance of the global cost gradient against the number of
    system qubits $n$, at hardware-efficient ansatz depths
    $\ell = 2$ to $5$.  Each point averages the per-parameter variance
    over five randomly chosen parameters from $200$ samples; error bars
    are $95\%$ confidence intervals.}
  \label{fig:bp_variance}
\end{figure*}

An
optimizer concentrates its evaluations where it has found descent, and it is the gradient
structure of those regions, not of the parameter space at large, that
governs whether optimization proceeds.
Figure~\ref{fig:bp_variance_trajectory} repeats the measurement with
$\bm{\theta}$ drawn from the points CMA-ES actually visits over $100$
generations. The variances are one to
three orders of magnitude above the uniform baseline at every size, and the apparent decay rate varies strongly with depth, in contrast to the uniform sampling case.

We emphasize that this is an observation about where the optimizer goes,
not a demonstration that its search strategy causes the effect: elevated
gradient variance along a trajectory is in part a consequence of
successful optimization as much as a cause of it.  What the measurement
does establish is that the regions CMA-ES occupies are atypical of the
landscape as a whole, in exactly the manner the narrow-gorge picture
of~\cite{arrasmith2022equivalence} anticipates, and that the exponential
suppression of variance of cost gradient is substantially
weaker along the path actually taken.

\subsubsection{Consequences for optimizer choice}

\begin{figure}[t]
  \centering
  \includegraphics[width=0.5\textwidth]{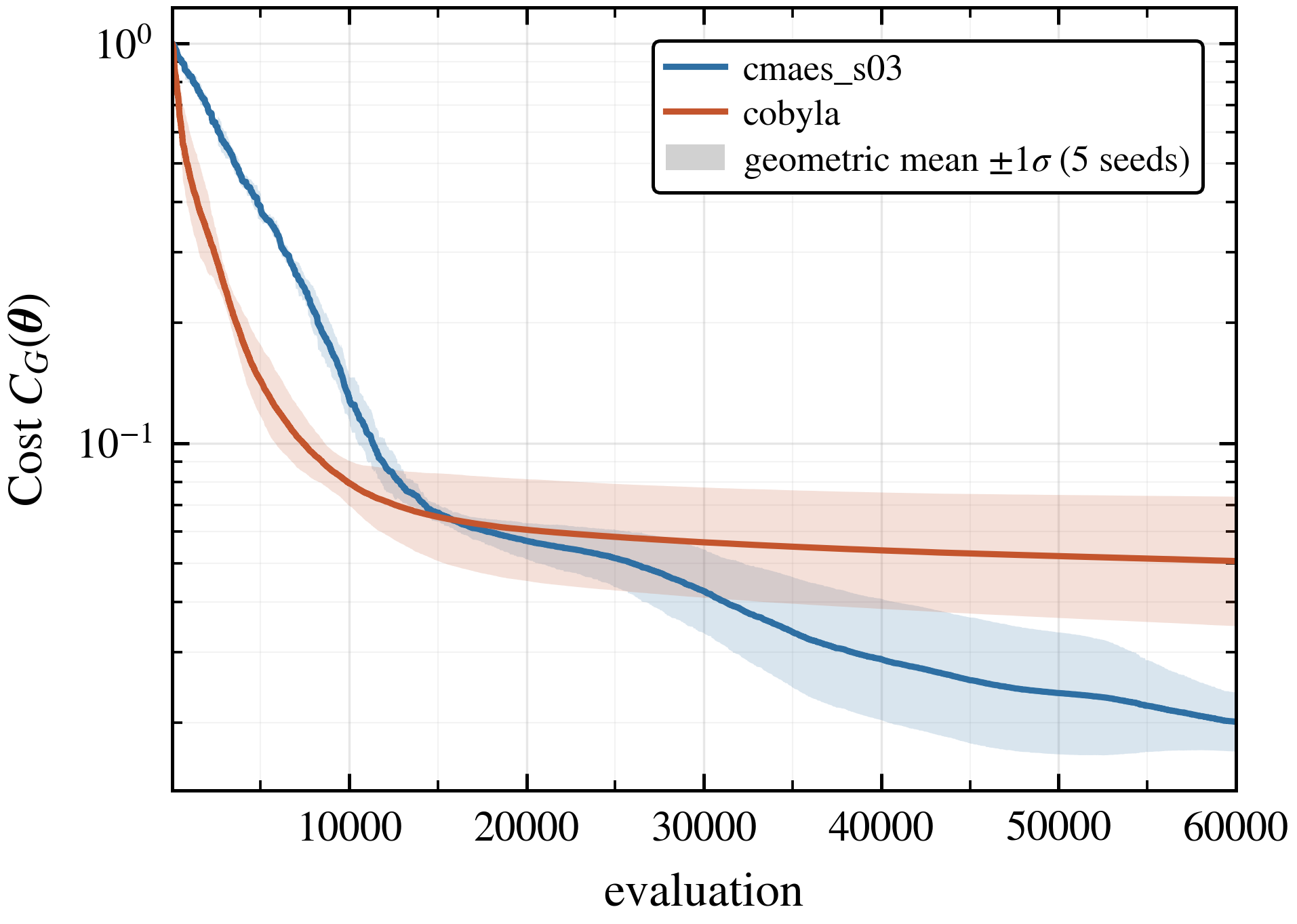}

    \caption{Cost function convergence for COBYLA and CMA-ES on the heat
  conduction benchmark ($N = 1024$) at equal budget: $4{,}000$ CMA-ES
  generations of population $15$ against $6 \times 10^{4}$ COBYLA
  iterations, i.e.\ $6 \times 10^{4}$ cost-function evaluations each.
  Solid lines are the geometric mean of the best-so-far cost across five
  seeds; bands are $\pm 1$ standard deviation in $\log_{10}$.}
  \label{fig:convergence_comparison}
\end{figure}

The practical question is whether this difference is visible in
convergence behaviour, and in Figure~\ref{fig:convergence_comparison}, the two optimizers are compared at equal cost: a
CMA-ES generation evaluates the cost function once per population
member, so $4{,}000$ generations at a population of $15$ corresponds to
the $6 \times 10^{4}$ evaluations allotted to COBYLA.

At $N = 1024$, COBYLA descends rapidly over the first few thousand
evaluations and then flattens, making
no further progress within its budget: its local linear model cannot identify productive descent directions once typical gradients are suppressed.  CMA-ES continues to reduce the cost throughout.

A further practical difference is that for the population within a CMA-ES
generation all cost evaluations are
independent and can be run in parallel.  COBYLA is inherently sequential,
each iteration depending on the outcome of the last.  This does not
alter the number of circuit calls, and on a single quantum
processor circuits execute serially in any case, but it is relevant
wherever multiple devices or simulator instances are available.

It is important to note that CMA-ES does not remove the barren plateau which is fundamental to the cost landscape.  It postpones the size at which stagnation sets
in, which is what the difference between the two panels of
Figure~\ref{fig:bp_variance} would lead one to expect.

\section{Conclusion and Future Work}
\label{sec:conclusion}

We have developed a variational quantum linear solver for the discrete
Poisson equation built on an exact block encoding of the Laplacian, and
characterized it on three benchmarks.

The encoding reproduces the target operator to double-precision
machine epsilon at every size from $N = 4$ to $N = 1024$, for Dirichlet
and Neumann boundaries alike.  Its post-selection success
probability decays as $N^{-0.57}$ and $N^{-0.94}$ for the two boundary
conditions, far more slowly than the $N^{-2}$ that would cancel the
advantage of reducing the cost function to a single circuit.  Against
the Pauli-LCU formulation, the difference is one of circuit count: the
Pauli-LCU VQLS requires $\Theta(N)$ circuits per cost evaluation,
against one in the block encoded VQLS.  Per circuit the LCU is cheaper at small sizes, but the ordering reverses beyond $N = 512$ in both gate count and depth, at a
fixed cost of three additional ancilla qubits.

Solution accuracy separates sharply between the two Dirichlet
benchmarks, and the reason is instructive.  The sinusoidal forcing is an
exact eigenvector of the discrete Laplacian, so its solution state
coincides with $|b\rangle$; the solver recovers it to arithmetic precision at
every size, which verifies the pipeline end to end but does not probe
its scaling.  For the Gaussian source, the infidelity remains at
machine precision through $N = 32$, then grows by roughly an order of
magnitude per grid size increase.

That degradation is attributable to the optimization.  Direct
measurement confirms a barren plateau: the gradient variance under
uniform sampling decays as $b^{-n}$,
statistically indistinguishable across ansatz depths $\ell = 2$ to $5$
and therefore driven by the growth of the Hilbert space rather than by
circuit depth.  Sampled instead along the CMA-ES trajectory, the
variance is one to three orders of magnitude higher at every size.  The
regions the optimizer occupies are thus atypical of the landscape as a
whole, which is consistent with its continuing to make progress at sizes
where COBYLA stalls, though we do not claim a causal mechanism.

Embedded in a projection-method Navier--Stokes loop, the solver
sustains a lid-driven cavity simulation at $Re = 100$ on a
$16 \times 16$ grid, tracking the classical trajectory in both
streamline structure and centreline profiles over ten time steps.

Several directions follow. On the encoding side, the construction
extends without modification to mixed boundary conditions and to three
or more spatial dimensions~\cite{boutot2026}, and block encodings have
been developed for non-uniform grids and non-Cartesian
meshes~\cite{Kharazi_2025}; both are needed before the approach can address
realistic geometries.  A quantitative comparison against compact,
non-Pauli LCU decompositions would sharpen the resource argument made in
Section~\ref{subsec:resources}. The block-encoding circuits used in this work are directly reusable as subroutines in Quantum Singular Value Transformation (QSVT)-based fault-tolerant quantum linear solvers~\cite{gilyen2019qsvt}, which is theoretically known to provide speedups over classical methods under well-defined conditions on matrix sparsity and condition number. 

On the optimization side, the empirical advantage of CMA-ES lacks a
theoretical account.  Understanding why a covariance-adapting search
locates and remains within the atypical regions identified here, and
whether that behaviour persists as $n$ grows, is the open question this
work raises most directly.  Related is the design of ans\"atze that
exploit the sparsity and Kronecker-sum structure of the Laplacian rather
than treating it as a generic operator.

The most promising path to practical quantum advantage and utility is to identify and efficiently obtain low-dimensional functionals of the solution rather than the full solution field, thereby circumventing the classical readout bottleneck. Identifying which quantities of engineering interest admit efficient quantum observable estimation is an important open problem that we intend to pursue in future work. Finally, all results reported here are from exact statevector
simulation.  This isolates the properties of the encoding and of the
optimization landscape from sampling and hardware noise, but leaves the
shot requirements and noise resilience of the method uncharacterized;
the success-probability scaling of
Section~\ref{subsec:resources} quantifies the former but does not
substitute for measuring it.  Establishing both is a prerequisite for
any hardware demonstration.

\begin{acknowledgments}

The authors thank Alexandre Boutot (University of Waterloo)
for his contributions to the block-encoding framework in a companion work~\cite{boutot2026}. The authors also thank Rajas K. Dalvi for his insights on the role of optimizers in delaying barren plateaus, and Kunal Garg for reviewing the manuscript. 
\end{acknowledgments}

\bibliography{references}
\end{document}